\documentclass[a4paper,12pt]{article}
\usepackage[utf8]{inputenc}
\usepackage{cancel}
\usepackage{ulem}
\usepackage{amsfonts}
\usepackage{amssymb}
\usepackage{graphicx}
\usepackage{amsmath}
\usepackage{enumerate}
\usepackage{mathtools}
\usepackage{subfig}
\usepackage{color}
\usepackage{tikz}\usetikzlibrary{calc}
\usepackage{float}
\usepackage{here}
\usepackage{cite}
\usepackage{mathrsfs}
\usepackage{float,epsfig}
\usepackage{dcolumn}

\usepackage{graphicx}
\usepackage{bm}
\usepackage{amsmath,amssymb,amsthm}
\usepackage[colorlinks=true,linkcolor=blue,citecolor=red]{hyperref}
\newcommand{\be}{\begin{equation}}
\newcommand{\ee}{\end{equation}}
\newcommand{\bea}{\setlength\arraycolsep{2pt} \begin{eqnarray}}
\newcommand{\eea}{\end{eqnarray}}

\def\0{{\sst{(0)}}}
\def\1{{\sst{(1)}}}
\def\2{{\sst{(2)}}}
\def\3{{\sst{(3)}}}
\def\4{{\sst{(4)}}}
\def\5{{\sst{(5)}}}
\def\6{{\sst{(6)}}}
\def\7{{\sst{(7)}}}
\def\8{{\sst{(8)}}}
\def\sst#1{{\scriptscriptstyle #1}}

\newcommand{\pgftextcircled}[1]{
    \setbox0=\hbox{#1}%
    \dimen0\wd0%
    \divide\dimen0 by 2%
    \begin{tikzpicture}[baseline=(a.base)]%
        \useasboundingbox (-\the\dimen0,0pt) rectangle (\the\dimen0,1pt);
        \node[circle,draw,outer sep=0pt,inner sep=0.1ex] (a) {#1};
    \end{tikzpicture}
}

\makeatletter \@addtoreset{equation}{section}

\usepackage{multirow}
\definecolor{lime}{HTML}{A6CE39}
\newcommand{\orcidicon}{%
    \begin{tikzpicture}
    \draw[lime, fill=lime] (0,0)
        circle [radius=0.16]
        node[white] {{\fontfamily{qag}\selectfont \tiny ID}};
    \draw[white, fill=white] (-0.0625,0.095)
        circle [radius=0.007];
    \end{tikzpicture}   \hspace{-2mm}
}
\newcommand\orcidAdil{{\href{https://orcid.org/0000-0001-7623-5541}{\orcidicon}}}
\newcommand\orcidHajar{{\href{https://orcid.org/0000-0001-9510-4248}{\orcidicon}}}              
\newcommand\orcidHasan{{\href{https://orcid.org/0000-0001-7408-0910}{\orcidicon}}}
\newcommand\orcidMohamed{{\href{https://orcid.org/0000-0003-1185-0062}{\orcidicon}}}

\begin{document}
%

\title{\normalsize
{\bf \Large	  Light  Behaviors  around    Black Holes    in M-theory  }}
\author{ \small   N.  Askour$^{1,2}$\footnote{askour2a@gmail.com}, A.  Belhaj,  \orcidAdil\!\! $^{3}$\footnote{a-belhaj@um5r.ac.ma},  H. Belmahi\orcidHajar\!\!$^{3}$\footnote{hajar\_belmahi@um5.ac.ma},  M. Benali\orcidMohamed\!\!$^{3}$\footnote{mohamed\_benali4@um5.ac.ma},    H. El Moumni\orcidHasan\!\!$^{4}$\thanks{h.elmoumni@uiz.ac.ma}, Y. Sekhmani$^{3}$\footnote{yassine\_sekhmani@um5.ac.ma} 
	\footnote{ Authors in alphabetical order.}
	\hspace*{-8pt} \\
	{\small $^1$ Department of Mathematics, Sultan Moulay Slimane University, Faculty of
Sciences and Technics }\\  {\small  B\'eni Mellal, BP 523, 23000, Morocco.}\\	
	{\small $^2$ Department of Mathematics, Mohammed V University in Rabat, Faculty of Sciences}\\  {\small  
Rabat, P.O. Box 1014, Morocco.}\\	
	{\small $^3$ D\'{e}partement de Physique, Equipe des Sciences de la mati\`ere et du rayonnement, ESMaR}\\
{\small   Facult\'e des Sciences, Universit\'e Mohammed V de Rabat, Rabat,  Morocco}\\
{\small $^{4}$  LPTHE, D\'{e}partement de Physique, Facult\'e des Sciences,   Universit\'e Ibn Zohr, Agadir, Morocco} } \maketitle

 \maketitle
\begin{abstract}
We study the deflection angle and the trajectory of the light rays around  black holes in M-theory scenarios. Using the Gauss-Bonnet theorem, we first compute and examine the deflection angle of the light rays near  four and seven dimensional AdS black holes obtained from the M-theory compactifications on  the real spheres on  $S^7$ and  $S^4$, respectively. We discuss the effect of the  M-theory brane number and the rotating parameter on such an optical quantity.   We then investigate the trajectories of the light rays using the equation of motion associated with $M2$ and $M5$-branes.

	\end{abstract}
\newpage

\tableofcontents

\section{Introduction}
Black hole physics has received a  remarkable interest encouraged by theoretical and observational findings \cite{SWH,sss}. These investigations have opened new gates, which could be exploited to understand gravity models from nontrivial theories including supergravity in higher dimensions \cite{emparan}. Concerning the observational results, the detection of the gravitational waves has been considered as a strong existence of the black hole objects \cite{A1}. These interesting observations are followed by the first image of the black hole shadows in  $M87^*$ galaxy provided by the Event Horizon Telescope (EHT) \cite{A2,A3,A4}. Motivated by such activities,   the thermodynamic and the optical properties of certain black holes have been extensively investigated using different methods based on analytic and numerical computations \cite{12,f2,D}. Various theories have been  exploited including the compactification of the superstrings and M-theory on the spherical internal compact manifolds. Inspired by the AdS/CFT correspondence,  it has been shown that the associated  black holes exhibit interesting thermodynamic behaviors. Precisely,  the  black holes on AdS geometries behave like Van der Wall fluid systems \cite{E1}. Interpreting the cosmological constant as a pressure, many phase transitions have been examined showing certain critical and universal aspects \cite{F,F1}. These thermodynamical investigations have been bridged with optical ones with the help of two essential concepts \cite{BW22,6,60}. The first treated concept is the deflection angle of the light rays near  black holes \cite{d1M,d2}. Concretely, Gibbons and Werner proposed a direct way to calculate this optical quantity by exploiting certain results of the  Gauss-Bonnet theorem using the optical metric analysis \cite{BW24}. Based on such a  method, the weak field limits of the deflection angle of the light rays around  various black holes have been investigated \cite{d3H,d4H,d5,d51,d6M}. Moreover,   the strong field deflection angle has received a large interest since it has been considered as a powerful tool to make contact with experimental measurements \cite{Boza1,Tien21}. The second one concerns the shadow. In four dimensions,  this optical aspect has been approached in terms of one-dimensional real curves \cite{K,BC,a40}. For the non-rotating solutions, these curves are identified with perfect circles. These geometrical configurations have been distorted by adding the rotating parameter \cite{Xa,J,RC}. In higher dimensions,  new geometrical configurations have been obtained where the shadows involve cardioid shapes. Concretely,  the shadow behaviors of the  five-dimensional (5D) black holes embedded in type IIB superstring/supergravity-inspired spacetimes have been dealt with by considering solutions with and without rotations \cite{H}. The geometrical properties have been analyzed in terms of the $D3$-brane number and the rotation parameter. It has been  shown that the shadows shapes are distorted by such parameters.   More precisely,  the size of the shadows decreases and gets deformed with the $M$-brane number.    Similar behaviors have been observed in four-dimensional black holes embedded in  M-theory inspired models \cite{B12}. It has been revealed that the $M2$-brane number can control the circular
shadow sizes. The geometrical behaviors are distorted for rotating solutions exhibiting cardioid shapes in certain moduli space regions.  Possible connections with observations (from Event Horizon Telescope or future devices) from a particular M-theory compactification have been proposed by deriving certain constraints on the $M2$-brane number in the light of the $M87*$ observational parameters. Generalizing the study to $d$-dimensional black holes surrounded by dark energy (DE), embedded in $D$ dimensional M-theory/superstring inspired models, the shadow geometries have been approached in terms of the $ M(d-2)$-brane number in the presence of DE \cite{ma}.

In this work,  we investigate the deflection angle and the trajectory of the light rays around the  black holes in M-theory spherical  compactifications. Using the Gauss-Bonnet theorem, we first compute and examine the deflection angle of the light rays by four and seven-dimensional AdS black holes obtained from the M-theory compactification on   the real spheres $S^7$ and  $S^4$, respectively.  We inspect the effect of the  $M$-theory brane number and the rotating parameter on such an optical quantity.   We then study the trajectories of the light rays using the equation of motion corresponding to  $M2$ and $M5$-branes.

This paper is organized as follows. In section 2, we generalize the deflection angle formalism for $d$-dimensional dealing with non-rotating AdS black holes using the Gauss-Bonnet theorem.  In section 3, we investigate the deflection angle of four-dimensional non-rotating and rotating AdS black holes by examining the $M2$-brane number effect in each case.  We extend the computation to seven-dimensional non-rotating black holes by providing a comparative study. In section 4,  we study the trajectory of the light rays around four and seven-dimensional AdS black holes in the M-theory compactification on   the real spheres $S^7$ and  $S^4$, respectively. The last section concerns concluding remarks.

\section{Deflection angle  formalism for  $d$-dimensional AdS black holes}
A close examination,    in the recent theoretical results,   shows many interesting optical properties  concerning black holes.  Certain investigations  have been extensively encouraged by EHT  collaboration using  different scenarios.  In particular,  the  shadows and the  light deflection of the light rays    have   been  examined providing interesting results in certain  four dimensional gravity theories.   For such reasons, we give the crucial formalism  used to compute  the deflection angle of the  light rays around   $d$-dimensional AdS black holes. Such an  optical quantity  has been  extensively investigated  using various methods and approximations \cite{BW18,BW27}.  The most powerful results have been derived from the Gauss-Bonnet theorem.  The calculation of the light deflection angle by   $d$-dimensional non-rotating AdS black holes can be established using the method developed in \cite{BW23}.   To start, we  consider  the ansatz metric
\begin{equation}
ds^{2}=-A(r)dt^{2}+B(r)dr^{2}+r^{2}d\Omega_{d-2}^{2}
 \end{equation}
where  $d\Omega_{d-2}^{2}\equiv d\theta^{2} +\prod_{k=1}^{d-3}\sin\theta \sin\phi_{k}$ represents the metric of the  $(d-2)$-real dimensional  unite sphere. $\theta$ and $\phi_k$ are the spheric local   coordinates.  $A(r)$  and $B(r)$ are  radial functions specified letter one.  The optical metric of the  light rays $\gamma_{ij}$ can be determined from the null condition, which  gives
\begin{equation}
dt^{2}=\gamma_{ij}dx^{i}dx^{j}.
\end{equation}
Such an optical metric can be expressed as 
\begin{equation}
dt^{2}= \frac{B(r)}{A(r)}dr^{2}+\frac{r^{2}}{A(r)}d\Omega_{d-2}^{2}
\end{equation}
providing a $(d-1)$dimensional Riemannian  space.  In this space,   we can calculate the deviation of the spatial curve by choosing an   equatorial plane  constrained  by   $\theta=\frac{\pi}{2}$. This plane is defined by the radial coordinate and one periodic coordinate of  $\{\phi_{k}\}$. The remaining ones should be fixed.   $\phi$  is considered as the only non fixed one. Before going ahead, certain conserved quantities are needed including the energy $E$ and the angular momentum   $L$. These conserved quantities are given by
\begin{eqnarray}
E&=& A(r)\frac{dt}{d\lambda}\\
{ L}&=& r^{2} \frac{d\phi}{d\lambda}
\end{eqnarray}
where $\lambda$ is the affine parameter.  They  can  be combined to define  the  impact parameter $b$ of  the motion  
\begin{eqnarray}
b =\frac{L}{E}.
\end{eqnarray} 
The computations  provide 
\begin{eqnarray}
b= \frac{r^{2}}{A(r)}\frac{d\phi}{d\lambda}.
\end{eqnarray} 
It turns out that certain vectors are needed, being the unit tangential vector of the light ray curve $K^{i}$ and the radial vector $R^{j}$.  They  are  defined as follows
\begin{eqnarray}
K^{i} &= &\frac{dx^{i}}{d\lambda}= \frac{b A(r)}{r^{2}}\left(\frac{dr}{d\phi},\underbrace{0,\dots,0}_\text{$d$-3},1\right),\\
R^{j}&=& \left( \frac{1}{\sqrt{\gamma_{rr}}},\underbrace{0,\dots,0}_\text{$d$-2}\right).
\end{eqnarray} 
Considering the plane  coordinated by $(r,\phi)$, the angle between the light rays and the radial direction can be extracted from the following relation
\begin{eqnarray}
\cos\Psi\equiv \gamma_{ij}K^iR^j.
\label{es}
\end{eqnarray}
The orbit equation can  be obtained from  the unity of the tangential vector  as
\begin{equation}
F_d(r)=\left( \frac{dr}{d\phi}\right)^{2} =\frac{r^{4}}{b^2 A(r)B(r)}-\frac{r^{2}}{B(r)}.
\label{es1}
\end{equation}
 In the plane   $(r,\phi)$, the calculations are reduced to the ones of the spherically symmetric black hole in four dimensions. Considering the observer ($R$) and the source ($S$)  at finite distances, one can define the deflection angle of the  light rays as follows
\begin{equation}
\Theta_d= \Psi_{R}-\Psi_{S}+\phi_{RS}
\end{equation}
where $\phi_{RS}$ is the separation angle. Putting $u=\frac{1}{r}$ in  Eq (\ref{es1}), this  optical angle  can be  rewritten as
\begin{eqnarray}
\phi_{RS} &= & \int^{u_0}_{u_S}\frac{1}{\sqrt{F_d(u)}}du +\int^{u_0}_{u_R}\frac{1}{\sqrt{F_d(u)}}du
\end{eqnarray}
where $u_{S}$ and $u_{R}$ are the inverse of the source and the observer distance from the black hole.  It is denoted that  $u_{0}$ is the inverse of the closest approach $r_{0}$ linked to the impact parameter via the constraint $F(u_{0})=0$. From the above formalism, we observe that  $A(r)$ and $B(r)$  functions encode all contribution effects including the dimension of the black hole space-time. To visualize such effects,  we investigate certain higher dimensional models.

\section{Light  deviating behaviors  of the black holes in  M-theory  compactification}
In this section, we study the deflection angle of the right rays around  black holes in  M-theory living in  11 dimensions.  At lower energy limits, this theory is described by 11-dimensional supergravity with $M2$ and $M5$-branes \cite{mtr3}.  Many black holes in such a theory have been constructed  \cite{mtr3,mtr1,mtr2}. Recently, the thermodynamics and the shadow optical properties have been investigated by considering   $D-$dimensional  M-theory inspired models where  the associated  $d$-dimensional AdS black holes have been embedded. It has been assumed that the AdS models are supposed to be dual versions of   $(d -1)$ dimensional CFT living on the associated boundaries, which could be modeled in terms of $N$ coincides  $(d -2)$ branes. It has been suggested that these AdS black holes can be obtained from the compactification of the   $D$-dimensional theories on the real spheres.  According to \cite{msm1},  the  associated near horizon geometries take the forms $AdS_{d}\times S^{d+k}$ where the dimension $D$ is constrained by
\begin{equation}
D=2d+k,
\end{equation}
with  $k$  is an integer indicating the dimension of    the 
 internal spherical spaces. In this way, the higher dimensional theories are indexed by a triplet $(D, d, k)$. For the non-rotating solutions, the line element of  the $d$-dimensional AdS black holes in such  a theory is given by  
\begin{eqnarray}\label{metric1}
		ds^2&=&-A(r)dt^2+\frac{dr^2}{A(r)}+r^2d^2\Omega_{d-2},		
	\end{eqnarray}
	where one has used the following metric function identification
	\begin{equation}
	B(r)=\frac{1}{A(r)}.
	\end{equation}
An inspection reveals that $A(r)$  is  a relevant metric function encoding data on the involved black hole  parameters.  In such $(D, d, k)$ models,  it takes the following  form 
	\begin{equation}
	A(r)=1-\frac{m}{r^{d-3}}+\frac{r^2}{L^2}_{AdS}.
	\end{equation}
In this equation,  $m$ is a physical parameter related to the black hole mass  via the relation 
\begin{equation}
		M=\frac{(d-2) \Omega_{d-2} m}{16 \pi G_d}.
		\label{ee4}
	\end{equation}
 where  $G_{d}$  and $\Omega_{d+k}$ are given by
 \begin{equation}
		G_{d}=\frac{\ell_{p}^{2(d-1)+k}}{\hbar \Omega_{d+K} L_{AdS}^{d+K}},\qquad  \Omega_{d+k}=\frac{2 \pi^{(d+k+1)/2}}{\Gamma(\frac{d+k+1}{2})}.
	\end{equation}
It is denoted   that $L_{AdS}$ is the radius  of the AdS space  related  to the number of  the $M-(d-2)$ branes as follows 
\begin{equation}
	L^{2(d-1)+k}_{AdS}=2^{-\left( \frac{d\left(4-d\right)+3}{2}  \right)} \, \pi^{7\left( k+2(d-5) \right)-4} \, N^{\frac{d-1}{2}} \ell_{p}^{ \, {2(d-1)+k}}.
	\label{ee5}	
	\end{equation}
	where  $\ell_{p}$ denotes the Planck length. To get  a concrete solution  derived from such M-theory  inspired models, the triplet  $(D,d,k)$ should be specified.   Phase transitions and shadow behaviors of such models have been investigated in \cite{B12,msm1,msm2}.   After that, an universal criticality of  the thermodynamic geometry of   the   $(D,d,k)$ models has been studied \cite{ma}.  
Motivated by such activities,  we examine the light behaviors around  AdS black holes for such   $(D,d,k)$ models.. We first deal with   the corresponding  deflection  angle of the  light rays.     For the sake of simplicities,  we will limit our analysis to triplets associated with M-theory.  We expect that the calculations could be adopted to  generic    $(D,d,k)$ models   \cite{msm3}.  However, the computations may  need  a deeper numerical computations.  At the end of this work, the light trajectories will be treated in section 4.

In the present section, however,  we  focus on the  deflection angle of the  light rays by  4-dimensional and 7-dimensional AdS black holes embedded in 11-dimensional M-theory associated with  the triplets  $(11,4,3)$  and $(11,7,-3)$, respectively.
\subsection{Light ray deflection in   the $(11,4,3)$  model}
\begin{figure}[!ht]
		\begin{center}
		\centering
			\begin{tabbing}
			\centering
			\hspace{8.cm}\=\kill
			\includegraphics[scale=0.8]{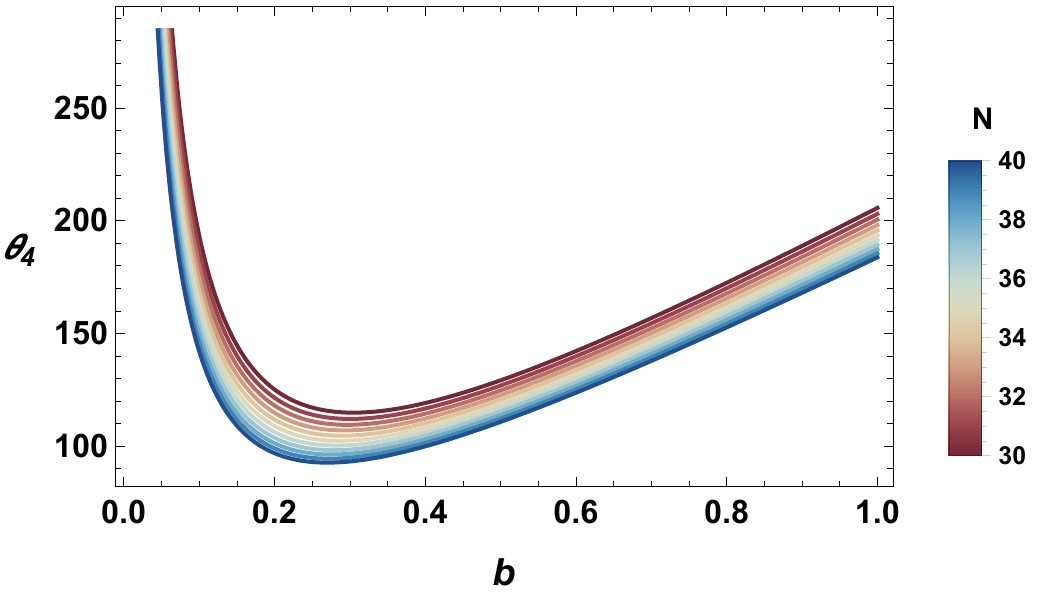} 
	\hspace{0.1cm}		\includegraphics[scale=0.8]{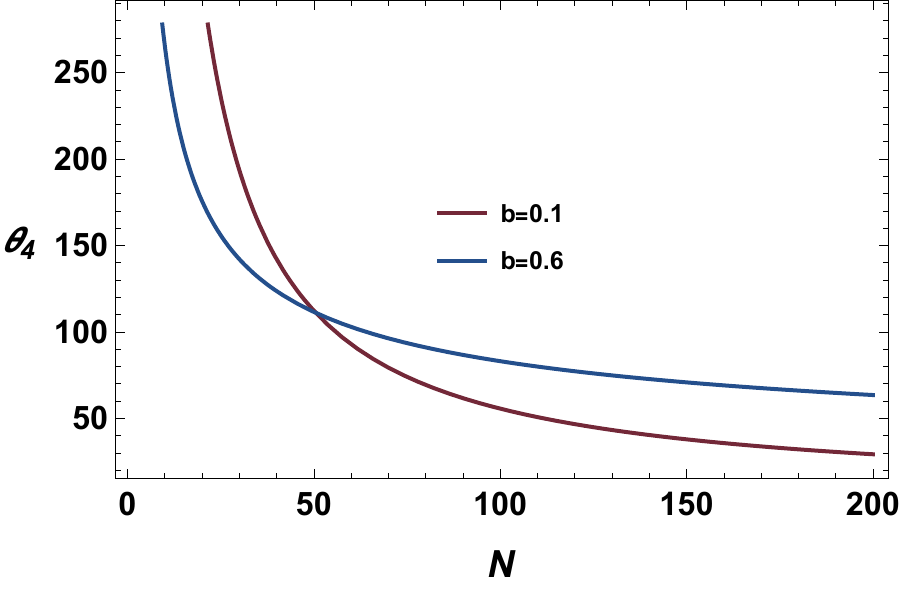}\\ 
		   \end{tabbing}
\caption{\footnotesize Right panel: Variation of the  deflection angle of  4-dimensional black holes in M-theory in terms of the impact parameter for different values of $N $. Lift panel: Variation of the  deflection angle of  4-dimensional black holes in M-theory in terms of the brane number for $b=0.1$ and $b=0.6$ }
\label{F1}
\end{center}
\end{figure}

We start by the non-rotating solutions corresponding to   the triplet $(11,4,3)$ obtianed from  M-theory  compactified  on  $S^{7}$ with $N$ coincide  $M2$-branes. Using Eq( \ref{ee4}) and Eq( \ref{ee5}), we get the metric function in terms of the  $M2$-brane  number   which reads as 
\begin{equation}
A(r)=1-\frac{192 \ 2^{\frac{1}{6}} \pi ^{\frac{2}{3}}\ell_{p}^2 M}{N^{\frac{7}{6}} r}+\frac{2^{\frac{1}{3}} r^2}{\pi ^{\frac{2}{3}} \ell_{p}^2 N^{\frac{1}{3}}},
\end{equation}
where one has used 
\begin{equation}
\label{L}
L_{AdS}=2^{-\frac{1}{6}} \pi^\frac{1}{3} N^{\frac{1}{3}} \ell_p,  \qquad m=\frac{192\times 2^{1/6} \,\pi ^{2/3}\,\ell_p^2 \, M}{N^{7/6}}. 
\end{equation}
The calculation will be expended to be in the leading order of $m$. The needed radial equation in four dimensions  is given by
\begin{equation}
\label{ma1}
F_{4}(u)=\frac{1}{b^2}-\frac{2^{\frac{1}{3}}}{\pi ^{\frac{2}{3}} \ell_{p}^2 N^{\frac{1}{3}}}-u^2+\frac{192 \ 2^{\frac{1}{6}}\pi ^{\frac{2}{3}} \ell_{p}^2 M }{N^{\frac{7}{6}}} u^3.
\end{equation}
Using the above equations,  the  angle  $\phi_{RS}$ can be written   as follows 
\begin{align}
&\phi_{RS}
=\left( \pi-\arcsin(bu_{R})-\arcsin(bu_{S})\right)  -\left(\frac{ u_R}{\sqrt{1-b^2 u_R^2}}+\frac{ u_S}{\sqrt{1-b^2 u_S^2}}\right)\frac{b^3}{(2 \pi )^{\frac{2}{3}}\ell _p^2 N^{\frac{1}{3}} }\notag\\
&+\left(\frac{2-b^2 u_R^2}{\sqrt{1-b^2 u_R^2}}+\frac{2-b^2 u_S^2}{\sqrt{1-b^2 u_S^2}}\right)\frac{96 \ 2^{\frac{1}{6}} \pi ^{\frac{2}{3}}  \ell _p^2 M}{b N^{\frac{7}{6}}} 
+\left(\frac{3 b^2 u_R^2-2}{\left(1-b^2 u_R^2\right){}^{3/2}}+\frac{3 b^2 u_S^2-2}{\left(1-b^2 u_S^2\right){}^{3/2}}\right)\frac{48 \sqrt{2} b M}{N^{3/2}}.
\label{E4}
\end{align}
 According to Eq(\ref{es}), the 
$\Psi_R - \Psi_S$ term can be expressed   as 
\begin{align}
&\Psi_R-\Psi_S
=\left( \arcsin(bu_{R})+\arcsin(bu_{S})-\pi\right)+\left(\frac{1}{u_R \sqrt{1-b^2 u_R^2}}+\frac{1}{u_S \sqrt{1-b^2 u_S^2}}\right)\frac{b}{(2 \pi )^{\frac{2}{3}}  \ell _p^2 N^{\frac{1}{3}}}\notag \\&-\left(\frac{u_R^2}{\sqrt{1-b^2 u_R^2}}+\frac{u_S^2}{\sqrt{1-b^2 u_S^2}}\right)\frac{96 \ 2^{\frac{1}{6}} \pi ^{\frac{2}{3}}\ell _p^2 b M }{N^{\frac{7}{6}}}+\left(\frac{1-2 b^2 u_R^2}{\left(1-b^2 u_R^2\right){}^{3/2}}+\frac{1-2 b^2 u_S^2}{\left(1-b^2 u_S^2\right){}^{3/2}}\right)\frac{48 \sqrt{2} b M}{N^{\frac{3}{2}}}.
\label{e2}
\end{align}
Combining the above equations, we can obtain  the deflection angle in terms of the $M2$-brane number $N$ and the  black hole mass.  this is found to be
\begin{align}
\Theta_{{4}}
=&
\left(\frac{\sqrt{1-b^{2}u_{R}^{2}}}{u_{R}}+\frac{\sqrt{1-b^{2}u_{S}^{2}}}{u_{S}}\right)\frac{b}{(2 \pi )^{\frac{2}{3}} \ell_{p} ^2{N}^{\frac{1}{3}}  }+
\left(\sqrt{1-b^{2}u_{R}^{2}}+\sqrt{1-b^{2}u_{S}^{2}}\right)\frac{192\pi ^{\frac{2}{3}} {2}^{\frac{1}{6}}\ell_{p} ^2  M }{b N^{\frac{7}{6}}}
\notag\\
& -\left(\frac{1}{\sqrt{1-b^{2}u_{R}^{2}}}
+\frac{1}{\sqrt{1-b^{2}u_{S}^{2}}}\right)\frac{48 \sqrt{2} b M}{N^{\frac{3}{2}}}. 
\label{E3}
\end{align}
It has been observed that the above expression diverges by taking the limits  $b u_S \to 0$ and $b u_R \to 0$. This is due to the fact that the spacetime is not asymptotically flat.  
Hence, the finite deflection angle of  the light rays by AdS  black holes from M-theory takes the form
\begin{equation}
\Theta_{{4}} \sim \frac{b }{(2 \pi )^{\frac{2}{3}}\ell_{p} ^2 {N}^{\frac{1}{3}} }\left(\frac{1}{{u_{R}}}+\frac{1}{{u_{S}}}\right)+\frac{384 \ 2^{\frac{1}{6}} \pi ^{\frac{2}{3}}\ell_{p} ^2 M }{b N^{\frac{7}{6}}}-\frac{96 \sqrt{2} b M}{N^{\frac{3}{2}}}.
\label{th1}
\end{equation}
In Fig.(\ref{F1}), we illustrate the  $M2$-brane  effect on such a deflection angle. In the left panel of this  figure, we present the variation of the deflection angle in terms of the impact parameter by taking different values of $N$.  Examining such  AdS black holes, we show that the deflection angle of the light rays decreases for small values of the impact parameter then it becomes an increasing function. It has been observed from the left and the right panels of Fig.(\ref{F1})  that the $M2$-brane number decreases the deflection angle. In the right panel,  we consider two values of the impact parameter. Plotting the deflection angle in terms of the  $M2$-brane number,  the two curves meet a particular point, where the deflection angle of the AdS space changes the behavior from a decreasing to an increasing function of the impact parameter.
A close examination shows that the behavior of the deflection angle of the AdS-Shwarzchild black holes for $M=1$ and $L=3$ is similar to the present one for $N\simeq147$. 

It seems possible to extend this  analysis of  the  deflection angle behaviors   in M-theory by introducing  the rotating parameter $a$.   According  \cite{msm4},   indeed, the metric  line element becomes  
\begin{equation}
ds^{2}=-\frac{\Delta_{r}}{W} \left(dt - \dfrac{a}{\Xi} \sin^2 \theta d\phi \right)^2 + W  \left(\frac{dr^2}{\Delta_{r}}+\frac{d\theta^2}{\Delta_\theta} \right) + \frac{\Delta_\theta \sin^2\theta}{W} \left( a dt -\frac{r^2+a^2}{\Xi}d\phi \right)^2.
\end{equation}
The  involved  terms  are given by
\begin{align}
 \Delta_{r}&=r^{2}-m r+a^{2}+\frac{r^{2}}{L^{2}}(r^{2}+a^{2}), ~~~~~~~~ \Delta_{\theta}=1-\frac{a^{2}}{L^{2}}\cos^{2}\theta, \\
 \Xi &= 1-\frac{a^{2}}{L^{2}},~~~~~~~~W=r^{2}+a^{2}\cos^{2}\theta
\end{align}
where $a$ is the rotating parameter. In this way, 
 the computation will be expanded to the first order of $m$ and $a$. To compute the $\Psi$ and $\phi_{RS}$ angles for such a rotating black hole in four dimensions,  we follow the method developed in \cite{BW18}, since we have the same metric form.  Considering the equatorial plane,  we can elaborate the orbit equation in terms of the $M2$-brane number  $N$ and the rotating parameter $a$.  In this way,  it  is found to be 
 \begin{equation}
 F_4(a,u)=\frac{1}{b^2}-u^2-\frac{\sqrt[3]{2}}{\pi ^{2/3} \sqrt[3]{N} \ell _p^2}+\frac{192 \sqrt[6]{2} \pi ^{2/3} M u^3 \ell _p^2}{N^{7/6}}+\frac{2 \sqrt[3]{2} a}{\pi ^{2/3} b^3 \sqrt[3]{N} u^2 \ell _p^2}-\frac{384 \sqrt[6]{2} \pi ^{2/3} a M u \ell _p^2}{b^3 N^{7/6}}+\frac{768 \sqrt{2} a M}{b^3 N^{3/2} u}.
   \end{equation}
 The longitudinal angle   can be  expressed as  follows
 \begin{eqnarray}
 &\phi_{RS}&=\left( \pi-\arcsin \left(b u_R\right)-\arcsin\left(b u_S\right)\right) -\left(\frac{u_R}{\sqrt{1-b^2 u_R^2}}+\frac{u_S}{\sqrt{1-b^2 u_S^2}}\right)\frac{b^3}{(2 \pi )^{2/3} \sqrt[3]{N} \ell _p^2}\notag \\&-&\left(\frac{1-2 b^2 u_R^2}{u_R \sqrt{1-b^2 u_R^2}}+\frac{1-2 b^2 u_S^2}{u_S \sqrt{1-b^2 u_S^2}}\right)\frac{\sqrt[3]{2} a}{\pi ^{2/3} \sqrt[3]{N} \ell _p^2}+\left(\frac{2-b^2 u_R^2}{\sqrt{1-b^2 u_R^2}}+\frac{2-b^2 u_S^2}{\sqrt{1-b^2 u_S^2}}\right)\frac{96 \sqrt[6]{2} \pi ^{2/3} M \ell _p^2}{b N^{7/6}}\notag\\&-&\left(\frac{1}{\sqrt{1-b^2 u_R^2}}+\frac{1}{\sqrt{1-b^2 u_S^2}}\right)\frac{192 \sqrt[6]{2} \pi ^{2/3} a M \ell _p^2}{b^2 N^{7/6}} -\left(\frac{2-3 b^2 u_R^2}{\left(1-b^2 u_R^2\right){}^{3/2}}+\frac{2-3 b^2 u_S^2}{\left(1-b^2 u_S^2\right){}^{3/2}}\right)\frac{48 \sqrt{2} b M}{N^{3/2}}\notag. 
   \end{eqnarray}
   Using the previous computations, we get  the  $\Psi$  angle 
   \begin{eqnarray}
   \Psi_{RS}&=&\left( \arcsin \left(b u_R\right)+\arcsin\left(b u_S\right)-\pi\right)+\left(\frac{1}{u_R \sqrt{1-b^2 u_R^2}}+\frac{1}{u_S \sqrt{1-b^2 u_S^2}}\right)\frac{b}{(2 \pi )^{2/3} \sqrt[3]{N} \ell _p^2}\notag\\&-&\left(\frac{u_R^2}{\sqrt{1-b^2 u_R^2}}+\frac{u_S^2}{\sqrt{1-b^2 u_S^2}}\right)\frac{96 \sqrt[6]{2} \pi ^{2/3} b M \ell _p^2}{N^{7/6}}+\left(\frac{1-2 b^2 u_R^2}{\left(1-b^2 u_R^2\right){}^{3/2}}+\frac{1-2 b^2 u_S^2}{\left(1-b^2 u_S^2\right){}^{3/2}}\right)\frac{48 \sqrt{2} b M}{N^{3/2}}\notag.
   \end{eqnarray}
   Combining the obtained  expressions,  we  can  obtain the deflection angle of four dimensional rotating AdS black holes  from M-theory  in terms of the involved parameters. The computations give 
   \begin{eqnarray}
   \Theta_4(a)&=&\left( \frac{1-b^2 u_R^2}{u_R \sqrt{1-b^2 u_R^2}}+\frac{1-b^2 u_S^2}{u_S \sqrt{1-b^2 u_S^2}}\right) \frac{b}{(2 \pi )^{2/3} \sqrt[3]{N} \ell _p^2}\notag\\&-&\left(\frac{1-2 b^2 u_R^2}{u_R \sqrt{1-b^2 u_R^2}}+\frac{1-2 b^2 u_S^2}{u_S \sqrt{1-b^2 u_S^2}}\right)\frac{\sqrt[3]{2} a}{\pi ^{2/3} \sqrt[3]{N} \ell _p^2}\notag\\&-&\left(\frac{1}{\sqrt{1-b^2 u_R^2}}+\frac{1}{\sqrt{1-b^2 u_S^2}}\right)\frac{84 \sqrt{2} b M}{N^{3/2}}+\left(\sqrt{1-b^2 u_R^2}+\sqrt{1-b^2 u_S^2}\right)\frac{192 \sqrt[6]{2} \pi ^{2/3} M \ell _p^2}{b N^{7/6}}\notag\\&-&\left(\frac{1}{\sqrt{1-b^2 u_R^2}}+\frac{1}{\sqrt{1-b^2 u_S^2}}\right)\frac{192 \sqrt[6]{2} \pi ^{2/3} a M \ell _p^2}{b^2 N^{7/6}}.
   \end{eqnarray}
 Taking a finite distance limit by  sending $b u_R$ and $b u_S$ to 0,  we   get the reduced expression of the deflection angle
   \begin{eqnarray}
  \Theta_4(a)&\sim&-\frac{96 \sqrt{2} b M}{N^{3/2}}+\left(\frac{1}{u_R}+\frac{1}{u_S}\right)\frac{b}{(2 \pi )^{2/3} \sqrt[3]{N} \ell _p^2}+\frac{384 \sqrt[6]{2} \pi ^{2/3} M \ell _p^2}{b N^{7/6}}-\left(\frac{1}{u_R}+\frac{1}{u_S}\right)\frac{\sqrt[3]{2} a}{\pi ^{2/3} \sqrt[3]{N} \ell _p^2}\notag\\ 
   &-&\frac{384 \sqrt[6]{2} \pi ^{2/3} a M \ell _p^2}{b^2 N^{7/6}}.
   \end{eqnarray}
  Putting $a=0$, we recover  the deflection angle of  four dimensional non-rotating black hole in M-theory given in  Eq.(\ref{th1}).
   \begin{figure}[!ht]
		\begin{center}
		\centering
			\begin{tabbing}
			\centering
			\hspace{8.cm}\=\kill
			\includegraphics[scale=0.75]{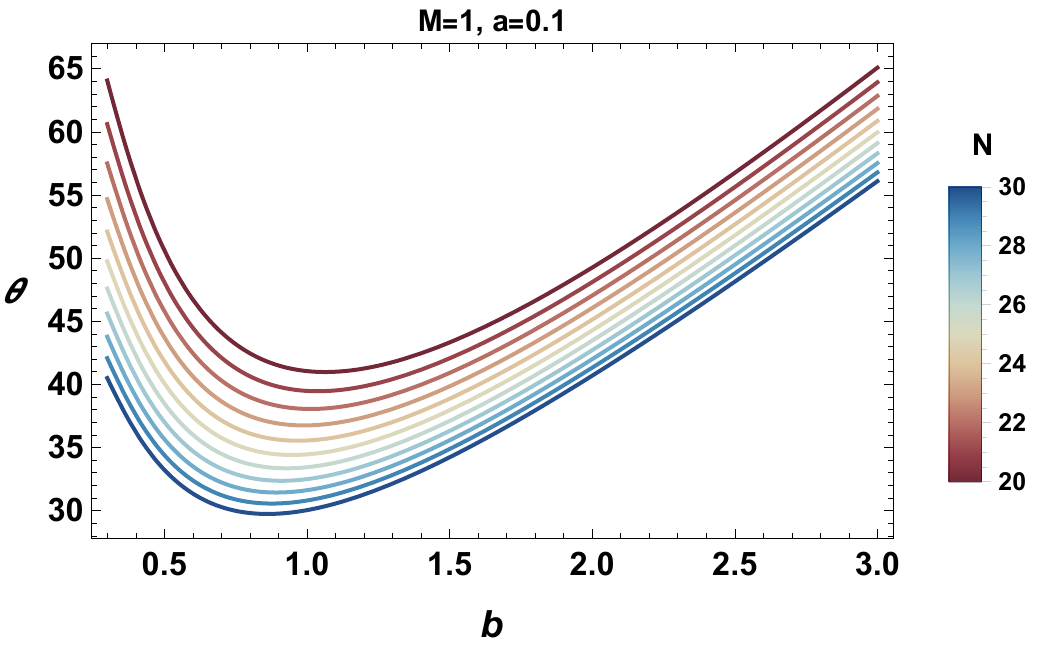} 
	\hspace{0.1cm}		\includegraphics[scale=0.75]{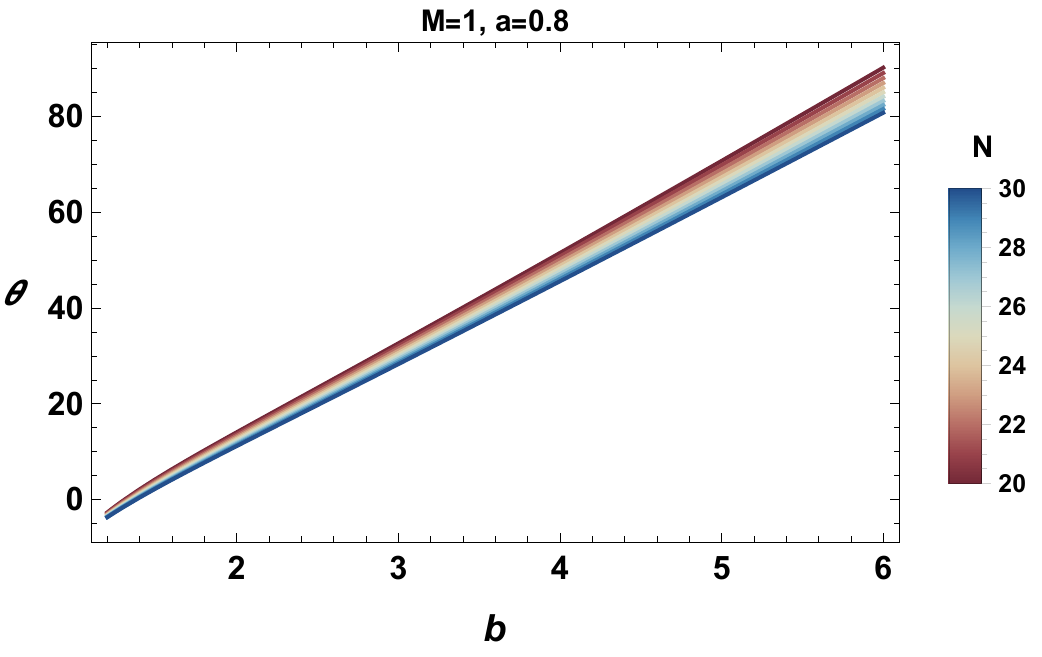}\\ 
	\includegraphics[scale=0.75]{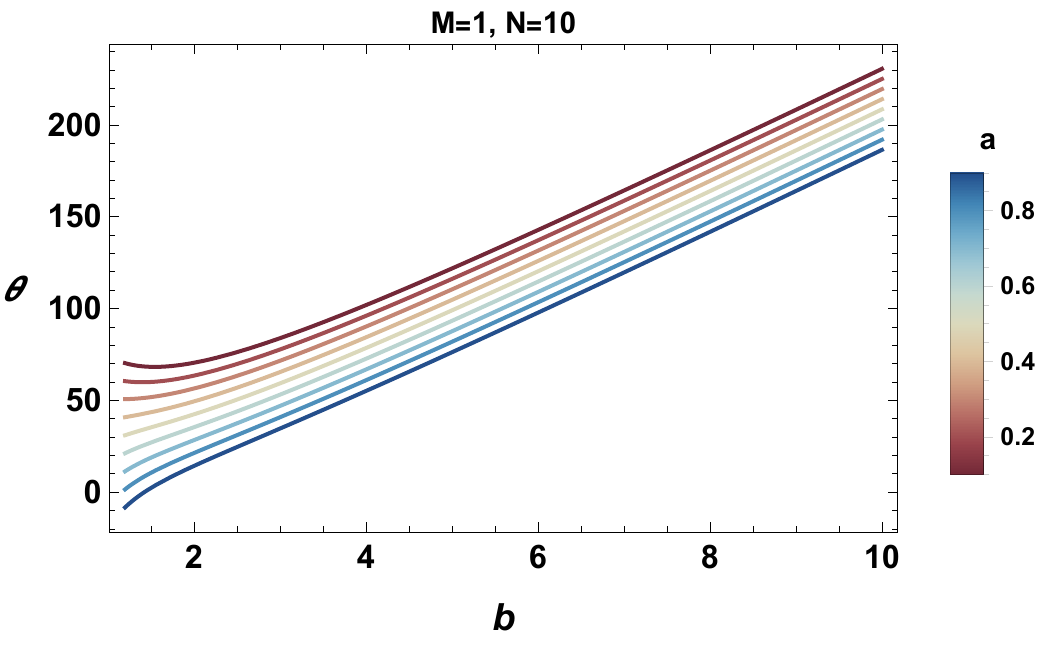} 
	\hspace{0.1cm}		\includegraphics[scale=0.75]{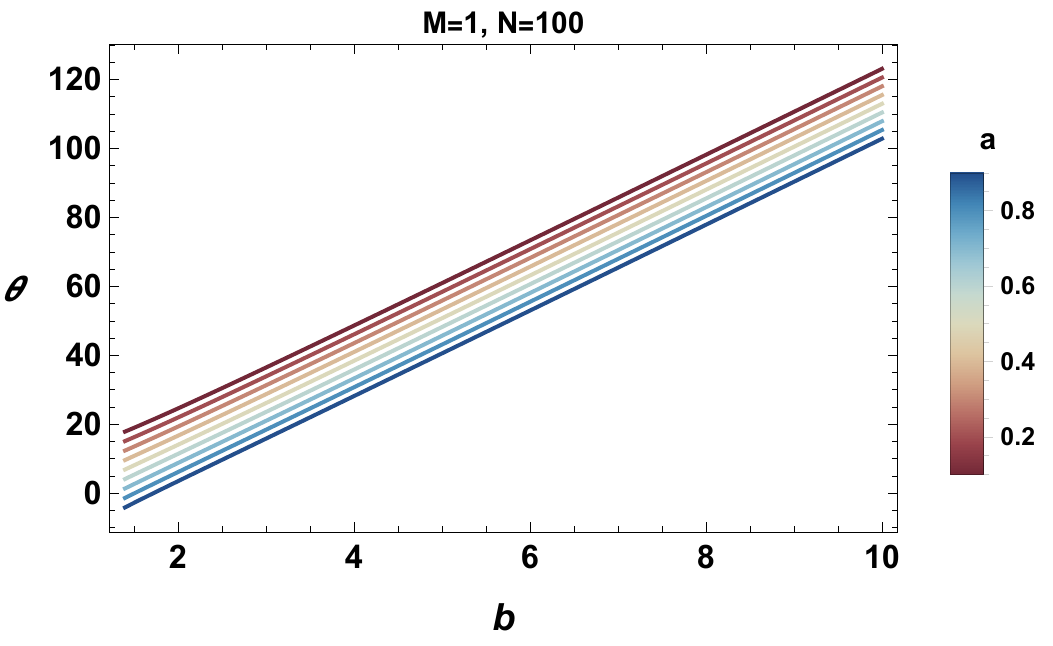}\\ 
		   \end{tabbing}
\caption{\footnotesize Variation of the deflection angle in terms of the impact parameter by varying $a$ and $N$.}
\label{F22}
\end{center}
\end{figure}

  To examine the effect of  the  $M2$-brane number  for  the  rotating black holes,  we vary the involved  parameters. Taking two values of the rotating parameter $a$, we plot  in  the top of the left and the right panels of Fig.(\ref{F22}) the variation of the deflection angle in terms of the impact parameter $b$ by varying   the  $M2$-brane number from 20 to 30. The  $M2$-brane number  parameter still decreases the deflection angle of the light rays.  In the left one, where we have a smaller value of $a$, the deflection angle of the light rays decreases for small values of $b$, and then it becomes an increasing function of the impact parameter $b$. In the right one,  where the contribution of the rotating parameter is relevant, the deflection angle is only an increasing function of $b$ without any minimum values.    In the bottom of Fig.(\ref{F22}),  we  take a small (left side) and a  large value (right side) of the $M2$-brane number and vary the rotating parameter  $a$ from 0.1 to 0.9. This shows that the rotating parameter $a$  still decreases the deflection angle as expected.  For generic values of the $M2$-brane number,  the deflection angle decreases by increasing the rotating parameter $a$. Near  $b=2$, the deflection angle behaviors depend on the $M2$-brane number. For generic values of $a$, the deflection angle increases by decreasing $N$. Similar behaviors have been obtained in the previous results.  The only difference around $b=2$ is the linear behavior for large  values of the  $M2$-brane number. It has been remarked that these optical behaviors could be related to the  AdS spacetime  backgrounds\cite{d3H,d4H,d6M}.

\subsection{Light deflection  behaviors  in   the $(11,7,-3)$  model}
Here, we consider  the 7-dimensional AdS black holes  by  considering the   triplet $(11,7,-3)$.   This model can be obtained from the compactification  of  M-theory on  four dimensional  real sphere $S^4$ in the presence of   $M5$-branes.  The  corresponding  metric function  in terms of the  $M5$-brane number  is expressed as follows
\begin{equation}
A(r)=1+\frac{r^2}{4 \pi ^{2/3} N^{2/3} \ell _p^2}-\frac{6 \pi ^{5/3} M \ell _p^3}{5 N^{4/3} r^4}.
\end{equation}
Using the orbital equation Eq(\ref{es1}), we obtain 
\begin{equation}
\label{ma2}
 F_7(u)=\frac{1}{b^2}-\frac{1}{4 \pi ^{2/3} N^{2/3} \ell _p^2}-u^2+\frac{6 \pi ^{5/3} M u^6 \ell _p^3}{5 N^{4/3}}.
\end{equation}
By the help of this  finding, we   get the $\phi_{RS}$ term
\begin{align}
\phi_{RS}
&=\left( \pi-\arcsin(bu_{R})-\arcsin(bu_{S})\right)- \left(\frac{u_R}{\sqrt{1-b^2 u_R^2}}+\frac{u_S}{\sqrt{1-b^2 u_S^2}}\right)\frac{b^3}{8 \pi ^{2/3} N^{2/3} \ell _p^2}\notag \\&+ \left(\frac{-2 b^5 u_R^5-5 b^3 u_R^3+15 b u_R}{\sqrt{1-b^2 u_R^2}}+\frac{-2 b^5 u_S^5-5 b^3 u_S^3+15 b u_S}{\sqrt{1-b^2 u_S^2}} \right)\frac{3 \pi ^{5/3} M \ell _p^3}{40 b^4 N^{4/3}}\notag \\&-\left(\frac{b u_R \left(3 b^4 u_R^4-20 b^2 u_R^2+15\right)}{\left(1-b^2 u_R^2\right){}^{3/2}}+\frac{b u_S \left(3 b^4 u_S^4-20 b^2 u_S^2+15\right)}{\left(1-b^2 u_S^2\right){}^{3/2}}\right)\frac{3 \pi  M \ell _p}{80 b^2 N^2}\notag \\&+\left( \pi-\arcsin(bu_{R})-\arcsin(bu_{S})\right)\left(\frac{9 \pi ^{5/3} M \ell _p^3}{8 b^4 N^{4/3}}-\frac{9 \pi  M \ell _p}{16 b^2 N^2}\right).
\label{ee2}
\end{align}
For the $\Psi$ part of these  black holes, we find  the following expression
\begin{align}
\Psi_R -\Psi_S
&=\left( \arcsin(bu_{R})+\arcsin(bu_{S})-\pi\right)+ \left(\frac{1}{u_R \sqrt{1-b^2 u_R^2}}+\frac{1}{u_S \sqrt{1-b^2 u_S^2}}\right)\frac{b}{8 \pi ^{2/3} N^{2/3} \ell _p^2}\notag \\&- \left(\frac{u_R^5}{\sqrt{1-b^2 u_R^2}}+\frac{u_S^5}{\sqrt{1-b^2 u_S^2}}\right)\frac{3 \pi ^{5/3} b M \ell _p^3}{5 N^{4/3}}\notag \\&-\left(\frac{u_R^3 \left(2 b^2 u_R^2-1\right)}{\left(1-b^2 u_R^2\right){}^{3/2}}+\frac{u_S^3 \left(2 b^2 u_S^2-1\right)}{\left(1-b^2 u_S^2\right){}^{3/2}}\right)\frac{3 \pi  b M \ell _p}{40 N^2}.
\label{ee1}
\end{align} 
Combining Eq(\ref{ee2}) and Eq(\ref{ee1}), we obtain    the deflection angle expression
\begin{align}
\Theta_{{7}}
=&
\left(\frac{\sqrt{1-b^{2}u_{R}^{2}}}{u_{R}}+\frac{\sqrt{1-b^{2}u_{S}^{2}}}{u_{S}}\right)\frac{b}{8 \pi ^{2/3} N^{2/3} \ell _p^2}\notag \\&-\left(\frac{b u_R \left(2 b^4 u_R^4+b^2 u_R^2-3\right)}{\sqrt{1-b^2 u_R^2}}+\frac{b u_S \left(2 b^4 u_S^4+b^2 u_S^2-3\right)}{\sqrt{1-b^2 u_S^2}}\right)\frac{3 \pi ^{5/3} M \ell _p^3}{40 b^4 N^{4/3}}\notag \\&+\left(\frac{b u_R \left(7 b^2 u_R^2-15\right)}{\sqrt{1-b^2 u_R^2}}+\frac{b u_S \left(7 b^2 u_S^2-15\right)}{\sqrt{1-b^2 u_S^2}}\right)\frac{3 \pi  M \ell _p}{80 b^2 N^2}\notag \\&+
\left( \pi-\arcsin(bu_{R})-\arcsin(bu_{S})\right)  \left(\frac{9 \pi ^{5/3} M \ell _p^3}{8 b^4 N^{4/3}}-\frac{9 \pi  M \ell _p}{16 b^2 N^2}\right).
\end{align}
Taking the finite distance limits by sending $b u_{S}$  and  $b u_{R}$ to zero, this deflection angle could be approximated   by the following form
\begin{align}
\Theta_{{7}}
&\sim 
\frac{9 \pi ^{8/3} M \ell _p^3}{8 b^4 N^{4/3}}-\frac{9 \pi ^2 M \ell _p}{16 b^2 N^2}+\frac{b }{8 \pi ^{2/3} N^{2/3} \ell _p^2}\left(\frac{1}{u_R}+\frac{1}{u_S}\right).
\end{align}
\begin{figure}[!ht]
		\begin{center}
		\centering
			\begin{tabbing}
			\centering
			\hspace{8.cm}\=\kill
			\includegraphics[scale=0.8]{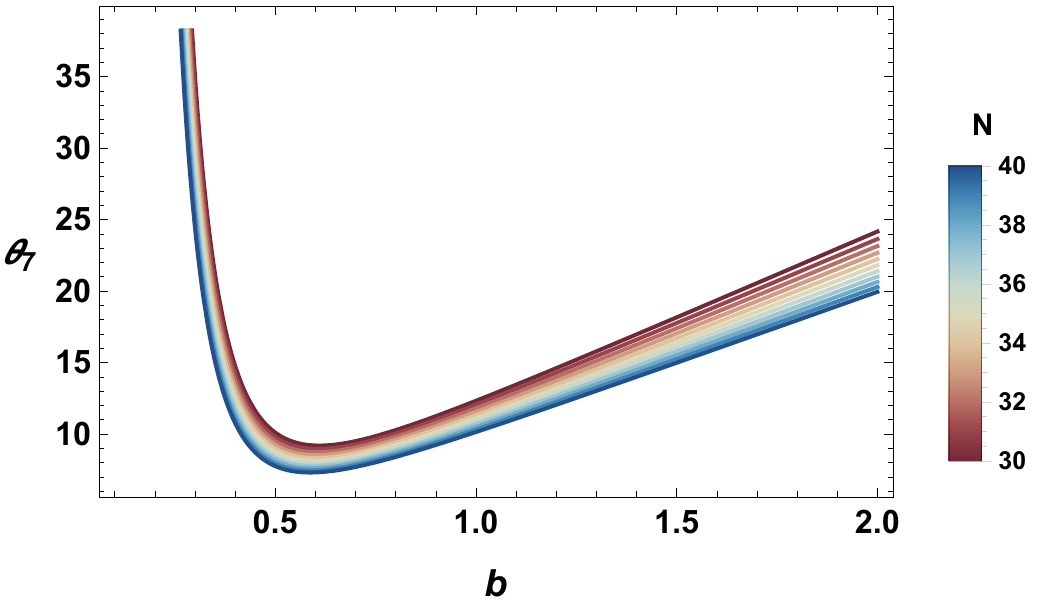} 
	\hspace{0.1cm}		\includegraphics[scale=0.8]{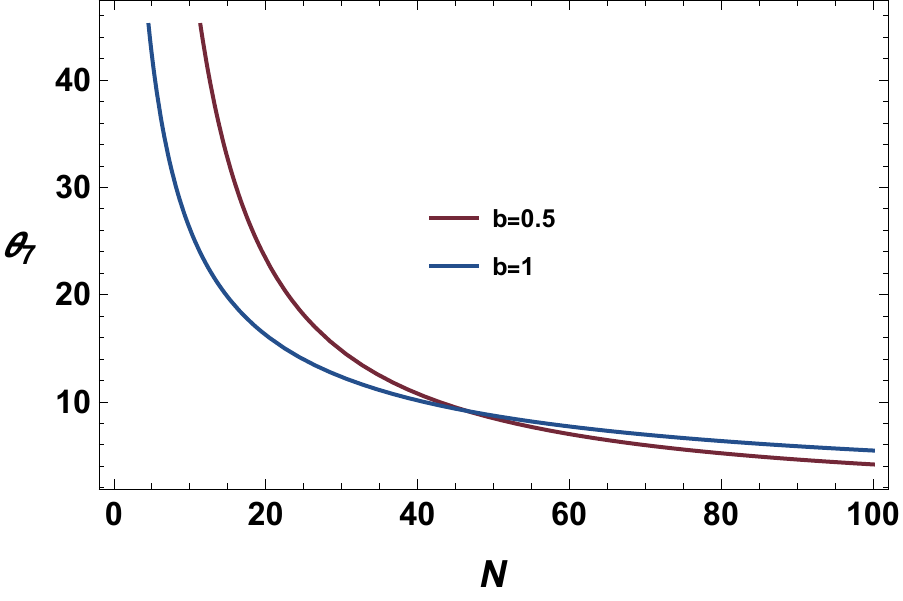}\\ 
	
		   \end{tabbing}
\caption{\footnotesize Right panel: Variation of the  deflection angle of  non rotating 7-dimensional black holes in M-theory in terms of the impact parameter for different values of $N $. Left panel: Variation of the  deflection angle of  7-dimensional black holes in M-theory in terms of the brane number for $b=0.5$ and $b=1$. }
\label{F2}
\end{center}
\end{figure}

To examine the associated behaviors, we plot in the left panel of Fig.(\ref{F2}) the variation of the deflection angle of the light rays by a 7-dimensional  black hole in M-theory in terms of the impact parameter by varying the  $M5$-brane number.  Similar to the four-dimensional case,  the deflection angle  of the light rays decreases for small values of the impact parameter to a critical value and then it  becomes an increasing function.  An examination of the right panel of Fig.(\ref{F2})  reveals that when we increase the number of $M2$-branes the deflection angle decreases. The  critical  transition behavior of the deflection angle from an increasing function to a decreasing one  is illustrated in the right panel of    Fig.(\ref{F2})  by the intersection point of the two curves.
  
\begin{figure}[!ht]
		\begin{center}

			\includegraphics[scale=0.8]{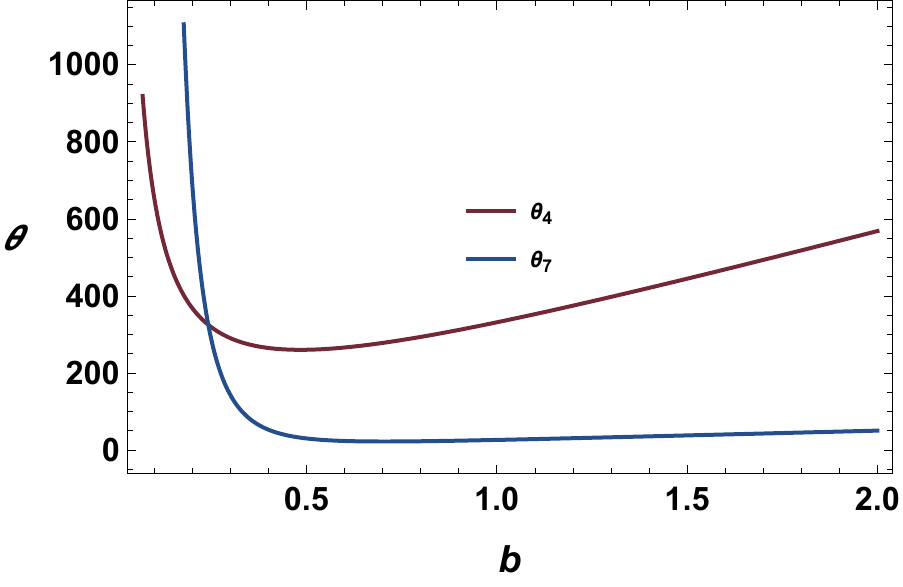} 

\caption{\footnotesize The variation of the deflection angle  in terms of the impact parameter  of four and seven dimensional non rotating black holes by taking $N=10$.}
\label{F333}
\end{center}
\end{figure}

 To compare the effect of the dimension on the variation of the deflection angle, we plot in the Fig.(\ref{F333}) the variation of the deflection angle in terms of the impact parameter for    $(11,4,3)$  and $(11,7,-3)$ models by fixing $N$.
For small values of the impact parameter, the four-dimensional black hole bends the light rays larger than the seven-dimensional black hole.  However, this behavior is inverted for large values of $b$. \\
Having discussed the behaviors   of the  deflection angle of the light rays casted by AdS black holes  in M-theory  in  the presence of $M2$ and $M5$-branes, we move to investigate the involved trajectories.   This has been motivated from   the fact that the light trajectories of the  black holes rely on the orbit  equation exploited in the deflection angle computations.   The study of  the  light trajectories  around the black holes for  $M2$ and $M5$-brane  models could  confirm  the behavior of the deflection angle  obtained in the previous subsections.

\section{ Light trajectories around  black holes in M-theory  }
In this section, we study the trajectory of the light rays by the black holes in M-theory scenarios.  Concretely,   we study the light trajectories near  the  AdS black holes of  ($D$, $d$, $k$) models  by varying  the  $M(d-2)$-brane number.   It is denoted that the light trajectories  around black holes can be generally approached  via  the   numerical computations adopted to the Eq.(\ref{ma1}) and Eq.(\ref{ma2}).    Concretely,  we can solve $\phi$ with respect to $u$  in order to depict  the behaviors of the light rays around  the involved black holes. 
 To establish  such   trajectories,   we need to identify    the regions  corresponding to   the light ray  trajectory possibilities.   These regions can be determined by  the help of the effective potential which is expressed as follows
 \begin{equation}
V^{eff}_{d}(r)=-\left( \frac{dr}{d\lambda}\right) ^{2}.
\end{equation}
This can be rewritten as 
 \begin{equation}
V^{eff}_{d}(r)=-F_{d}(r)\left( \frac{b A(r)}{r^2}\right)^2.
\end{equation}
 For simplicity reasons,   we restrict ourselves to the special  models    embedded in 11-dimensional supergravity limits of M-theory.
 \subsection{Trajectories of the light rays  in the  $(11,4,3)$  model}
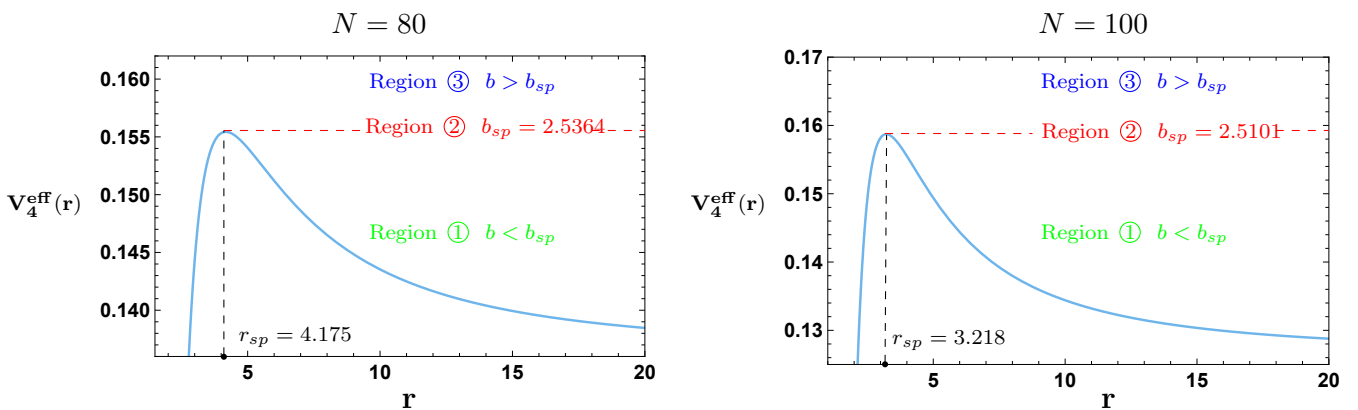
\begin{figure*}[!ht]
		\begin{center}
		\begin{tikzpicture}[scale=0.2,text centered]
		\hspace{-1.2 cm}
\hspace{0.1 cm}{\node[] at (-40,1){\small  \includegraphics[scale=0.8]{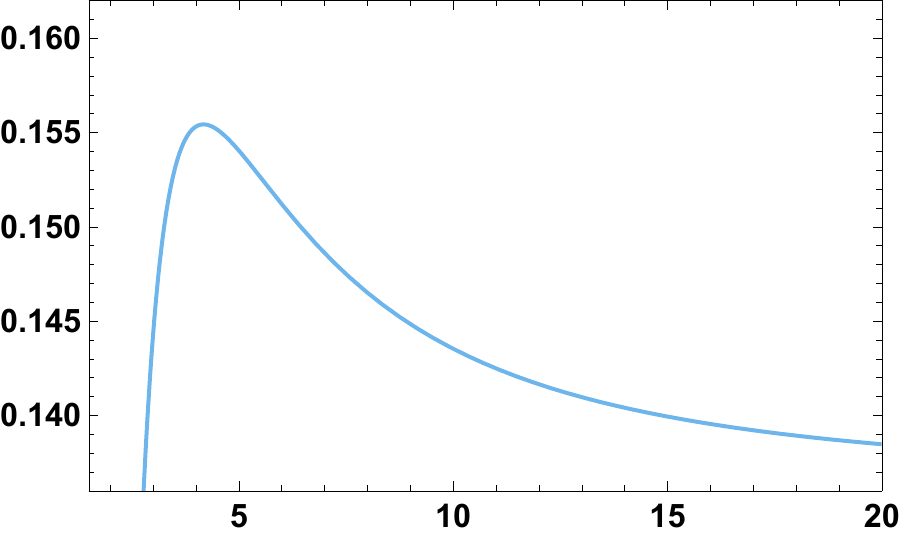}};
\node[] at (5,1){\small  \includegraphics[scale=0.8]{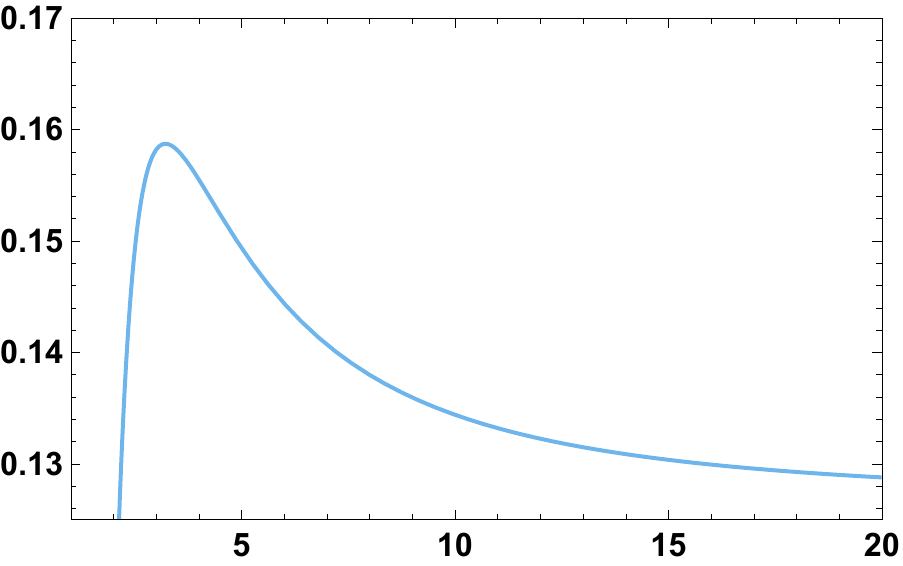}};}
\vspace{4pt}
\draw[dashed] (-6.7,-8.5) -- (-6.6,6.6);
\draw[dashed,color=red] (-6.7,6.7) -- (3,6.7);
\draw[dashed,color=red] (19,6.9) -- (22.5,6.9);
\draw[dashed] (-50.2,-8) -- (-50.2,6.9);
\draw[dashed,color=red] (-50.2,6.9) -- (-41,6.9);
\draw[dashed,color=red] (-26,6.9) -- (-22.5,6.9);
\node[color=black] at (-40,14){\small {$N = 80 $}};
\node[color=black] at (7,14){\small {$N= 100 $}};
\node[color=black] at (-34.5,10){\scriptsize {\color{blue}{Region \let\textcircled=\pgftextcircled\textcircled{3}}\; $b>b_{sp}$}};
\node[color=black] at (-33,7){\scriptsize {\color{red}{Region \let\textcircled=\pgftextcircled\textcircled{2}}\; $b_{sp}=2.5364$}};
\node[color=black] at (-34.5,0){\scriptsize {\color{green}{Region \let\textcircled=\pgftextcircled\textcircled{1}}\; $b<b_{sp}$}};
\node[color=black] at (9.7,10){\scriptsize {\color{blue}{Region \let\textcircled=\pgftextcircled\textcircled{3}}\; $b>b_{sp}$}};
\node[color=black] at (11.5,6.7){\scriptsize {\color{red}{Region \let\textcircled=\pgftextcircled\textcircled{2}}\; $b_{sp}=2.5101$}};
\node[color=black] at (9.7,0){\scriptsize {\color{green}{Region \let\textcircled=\pgftextcircled\textcircled{1}}\; $b<b_{sp}$}};
\fill (-6.7,-8.6) circle (0.2);
\node[color=black] at (-2.5,-7){\scriptsize {$r_{sp}=3.218$}};
\fill (-50.2,-8.1) circle (0.2);
\node[color=black] at (-45.5,-6.8){\scriptsize {$r_{sp}=4.175$}};
\node[color=black] at (-62,2){\scriptsize  $\mathbf{V_{4}^{eff}(r)}$};
\node[color=black] at (-17,2){\scriptsize  $\mathbf{V_{4}^{eff}(r)}$};
\node[color=black] at (-38,-11){$\mathbf{r}$};
\node[color=black] at (8,-11){$\mathbf{r}$};
\end{tikzpicture}	
\caption{{ \footnotesize  The effective potential behaviors  of  4-dimensional  AdS black holes embedded
in 11-dimensional M-theory by varying taking to values of   the brane number.}}
\label{aa1b}
\end{center}
\end{figure*}

 We start by considering   the  $(11,4,3)$  model developed in \cite{B12,ma}.  Evincing  the  rotating parameter, the four dimensional effective potential, in the presence of  the $M2$-branes,   takes the following form
\begin{equation}
\label{ }
V^{eff}_{4}(r)=\frac{{ L}^2}{r^2}\left(1-\frac{192 \ 2^{\frac{1}{6}} \pi ^{\frac{2}{3}}\ell_{p}^2 M}{N^{\frac{7}{6}} r}+\frac{2^{\frac{1}{3}} r^2}{\pi ^{\frac{2}{3}} \ell_{p}^2 N^{\frac{1}{3}}}\right)-E^2.
\end{equation}
 This  effective potential will be illustrated  as  a function of the radial coordinate $r$ for different values of  the   $M2$-brane number $N$ by taking  $\ell_p=1$ and $M=1$.   According to \cite{B12}, the maximum value of the shadow radius  corresponds to $N=80$.   In  Fig.(\ref{aa1b}),   we plot  such    a potential for  two  $M2$-brane number   values being  $N=100$ and $N=80$. 
 
 This potential increases and reaches a maximum at the photon sphere associated with    $b_{sp}$   which represents   the impact parameter of   the spinning  light rays around  the black holes.  This  verifies the following constraint 
\begin{equation}
\label{p1}
V^{eff}(r_{sp})=\frac{1}{b^2_{sp}}.
\end{equation} 
It has been found that the  two values of  the M2-brane number $N=80, 100$  provide two   impact parameter values  $b_{sp}=2.5364$ and   $b_{sp}= 2.5101$ corresponding  to the  photon sphere radius $r_{sp}=4.175$ and  $r_{sp}=3.218$, respectively as   shown   in  Fig.(\ref{aa1b}). Indeed, 
 the associated   impact parameter and the photon sphere radius  decrease by increasing the M2-brane number.   It has been remarked  that the impact parameter value  $b_{sp}$ provides    trajectories  of the light rays  in three  different regions. These  regions are denoted by region \let\textcircled=\pgftextcircled\textcircled{1}, region \let\textcircled=\pgftextcircled\textcircled{2} and region \let\textcircled=\pgftextcircled\textcircled{3} corresponding to $b<b_{sp}$, $b=b_{sp}$ and $b>b_{sp}$, respectively.
 
\begin{figure*}[!ht]
		\begin{center}
		\begin{tikzpicture}[scale=0.2,text centered]
		\hspace{-0.5 cm}
\node[] at (-30,1){\small  \includegraphics[scale=0.50]{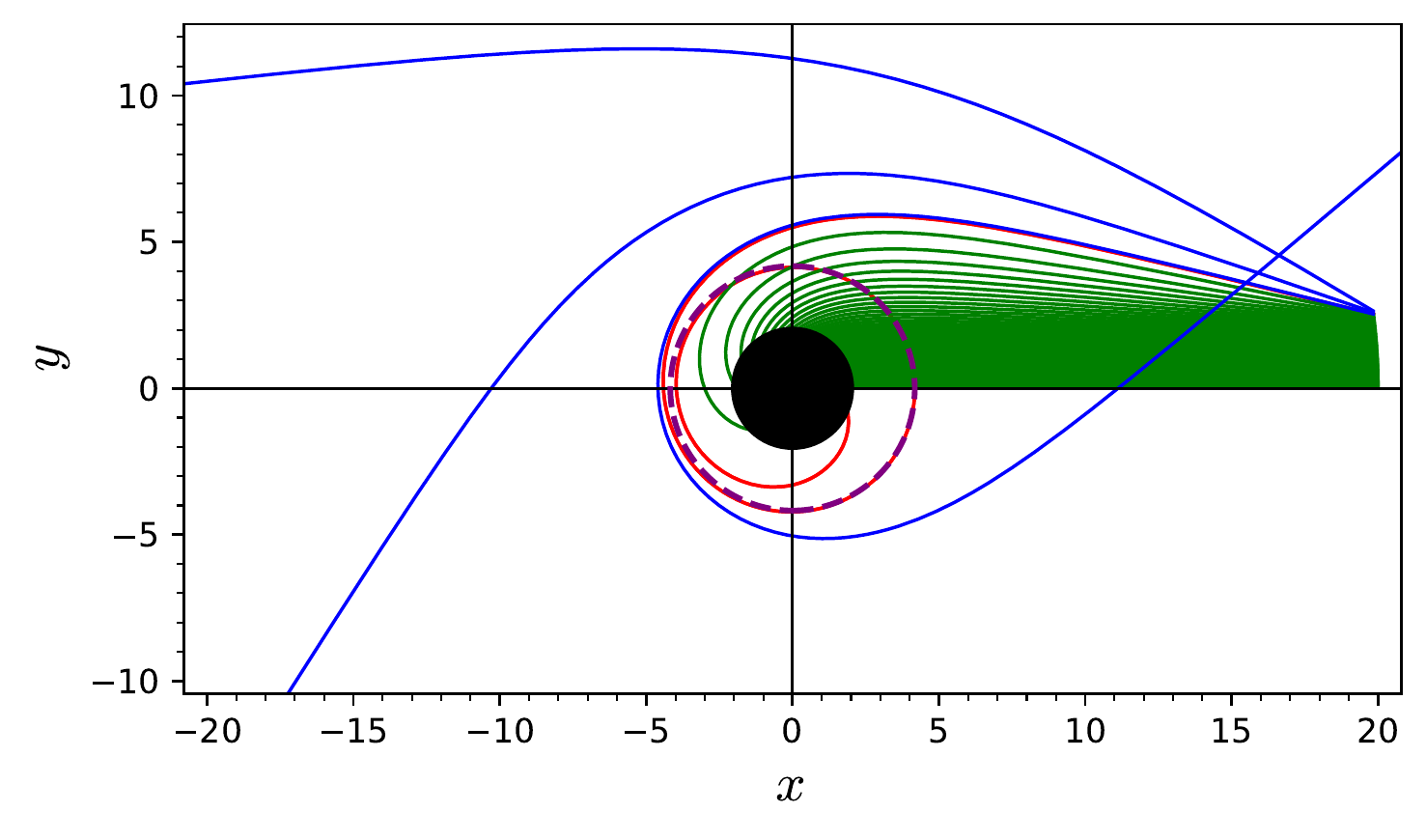}};
\node[] at (10,1){\small  \includegraphics[scale=0.50]{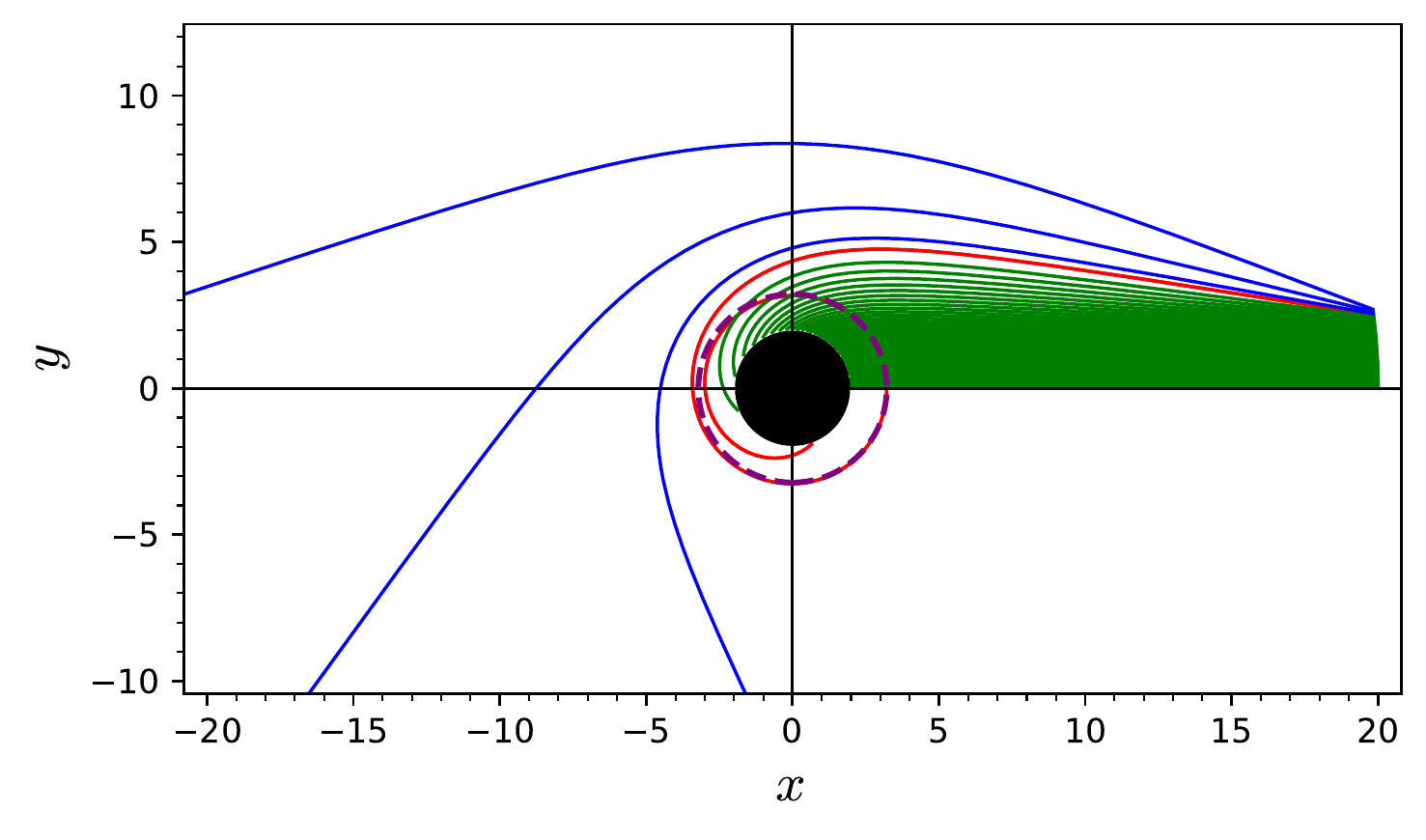}};
\node[color=black] at (-27.8,13.5) {$N=80$};
\node[color=black] at (12.8,13.5) {$N=100$};
\end{tikzpicture}	
\caption{ \footnotesize The trajectories of the light ray for different values of  $M2$-brane number. The black and the dashed red circle is the horizon and the photon sphere of the $M2$-brane, respectively.}
\label{aa}
\end{center}
\end{figure*}

In the first region  \let\textcircled=\pgftextcircled\textcircled{1},  the light ray falls  into the black hole due to the values of  the impact parameter lower to $b_{sp}$.  In the third  region \let\textcircled=\pgftextcircled\textcircled{3},  however, the light ray near the black hole can be reflected back.  In the  second  region   \let\textcircled=\pgftextcircled\textcircled{2},  however,   the light ray comes into the photon sphere making an infinite number of turns around the black  hole  due to the a non vanishing  angular velocity. The associated orbit is circular and unstable. To illustrate these regions,  we plot in Fig.(\ref{aa}) the trajectories of the light rays in the polar coordinates $(r,\phi)$ for different  values of the $M2$-brane number  $N$. To analyze the effect of the $M2$-brane  on the  light ray trajectories,  we vary the impact parameter $b$ by using   the step between  two values of   the impact parameter as 1/20 for all light rays.
A close examination  reveals  that the horizon and the sphere photon radius decrease by increasing the $M2$-brane number.  However,  the variation of the impact parameter $b_{sp}$  is small by varying the $M2$-brane number,  $N$.  It has been observed that the region  \let\textcircled=\pgftextcircled\textcircled{1}, \let\textcircled=\pgftextcircled\textcircled{2} and \let\textcircled=\pgftextcircled\textcircled{3}  are  the same for different  values of  the $M2$-brane number. Taking two values of $N$,  we observe that the  distance between  two   light rays increases for  an impact parameter value  closer and bigger to  $b_{sp}$  for all regions. Indeed, this distance decreases by increasing $N$.   This distinction  comes from the values of the    angular velocity.   More analysis shows  that the  reflected  of  the light ray is more intense by decreasing the  $M2$-brane number.
Comparing  this result  with many works concerning the trivial solution\cite{a37,a38,a39}, we observe a different behavior. First, $b_{sp}$   is almost the same by varying the  $N$.   However,  $r_{sp}$  decreases  by increasing   the  $M2$-brane number. This is completely different  than the  previous  results.  Second,  for small values of the  impact parameter, we  remark that the light ray falls  into the black hole by   keeping  the parallel trajectories with the  $r$-axis. However, for values close to $b_{sp}$  the light ray falls into the black hole without keeping  the parallel trajectory with  the $r$-axis. Indeed, the parallel trajectory is replaced by a critical angle between the light ray and the  $r$-axis. Fixing the $M2$-brane number, this angle increases for values near to  $b_{sp}$ (or bigger). Varying $N$,  it increases by decreasing the  $M2$-brane number.   This shows that such a  $M2$-brane number can be considered as a relevant quantity modifying the light ray behaviors near a black hole in M-theory compactifications.  This distinction comes from the geometry of the black holes in M-theory with brane  backgrounds. 
 \subsection{Light  trajectories  in the $(11,7,-3)$  model}
\begin{figure*}[!ht]
		\begin{center}
		\begin{tikzpicture}[scale=0.2,text centered]
		\hspace{-1.2 cm}
\hspace{0.1 cm}{\node[] at (-40,1){\small  \includegraphics[scale=0.8]{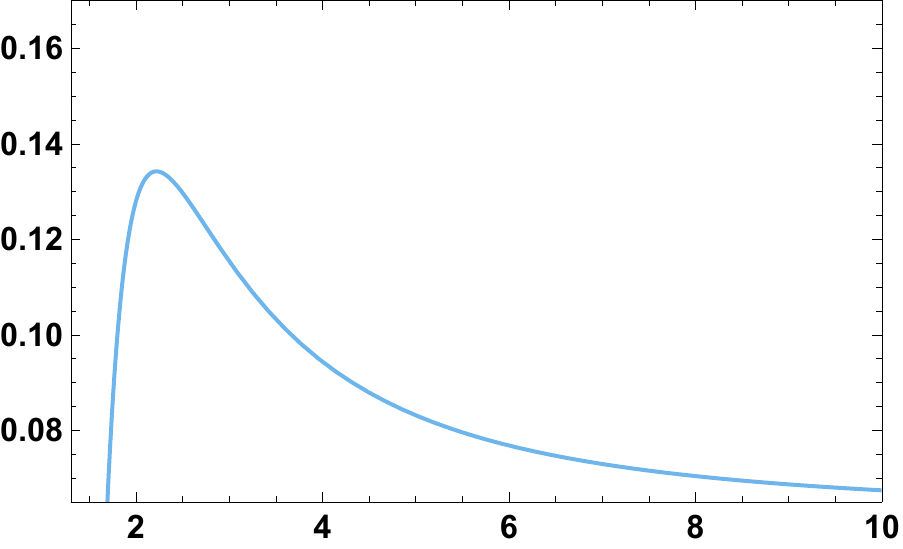}};
\node[] at (5,1){\small  \includegraphics[scale=0.8]{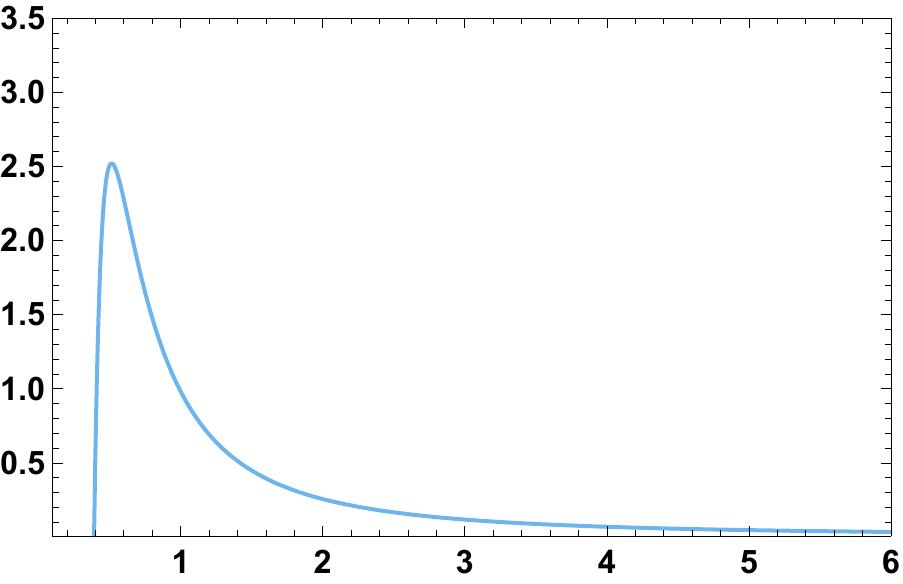}};}
\vspace{4pt}
\draw[dashed] (-8.7,-8.5) -- (-8.7,6.3);
\draw[dashed,color=red] (-8.7,6.3) -- (3,6.3);
\draw[dashed,color=red] (19.5,6.3) -- (22.5,6.3);

\draw[dashed] (-52,-8) -- (-52,5.3);
\draw[dashed,color=red] (-52,5.3) -- (-41,5.3);
\draw[dashed,color=red] (-26,5.3) -- (-22.5,5.3);
\node[color=black] at (-40,14){\small {$N = 1 $}};
\node[color=black] at (7,14){\small {$N= 80 $}};
\node[color=black] at (-34.5,10){\scriptsize {\color{blue}{Region \let\textcircled=\pgftextcircled\textcircled{3}}\; $b>b_{sp}$}};
\node[color=black] at (-33,5.5){\scriptsize {\color{red}{Region \let\textcircled=\pgftextcircled\textcircled{2}}\; $b_{sp}=2.7180$}};
\node[color=black] at (-34.5,0){\scriptsize {\color{green}{Region \let\textcircled=\pgftextcircled\textcircled{1}}\; $b<b_{sp}$}};
\node[color=black] at (9.7,10){\scriptsize {\color{blue}{Region \let\textcircled=\pgftextcircled\textcircled{3}}\; $b>b_{sp}$}};
\node[color=black] at (11.5,6.3){\scriptsize {\color{red}{Region \let\textcircled=\pgftextcircled\textcircled{2}}\; $b_{sp}=0.6308$}};
\node[color=black] at (9.7,0){\scriptsize {\color{green}{Region \let\textcircled=\pgftextcircled\textcircled{1}}\; $b<b_{sp}$}};
\fill (-8.7,-8.9) circle (0.2);
\node[color=black] at (-4.5,-8){\scriptsize {$r_{sp}=0.515$}};
\fill (-52,-8.2) circle (0.2);
\node[color=black] at (-47.5,-6.8){\scriptsize {$r_{sp}=2.219$}};
\node[color=black] at (-62,2){\scriptsize  $\mathbf{V_{7}^{eff}(r)}$};
\node[color=black] at (-17,2){\scriptsize  $\mathbf{V_{7}^{eff}(r)}$};
\node[color=black] at (-38,-11){$\mathbf{r}$};
\node[color=black] at (8,-11){$\mathbf{r}$};
\end{tikzpicture}	
\caption{{ \footnotesize  The effective potential behaviors  of  7-dimensional  AdS black holes embedded
in 11-dimensional M-theory by varying taking to values of   the brane number.}}
\label{aa2b}
\end{center}
\end{figure*}
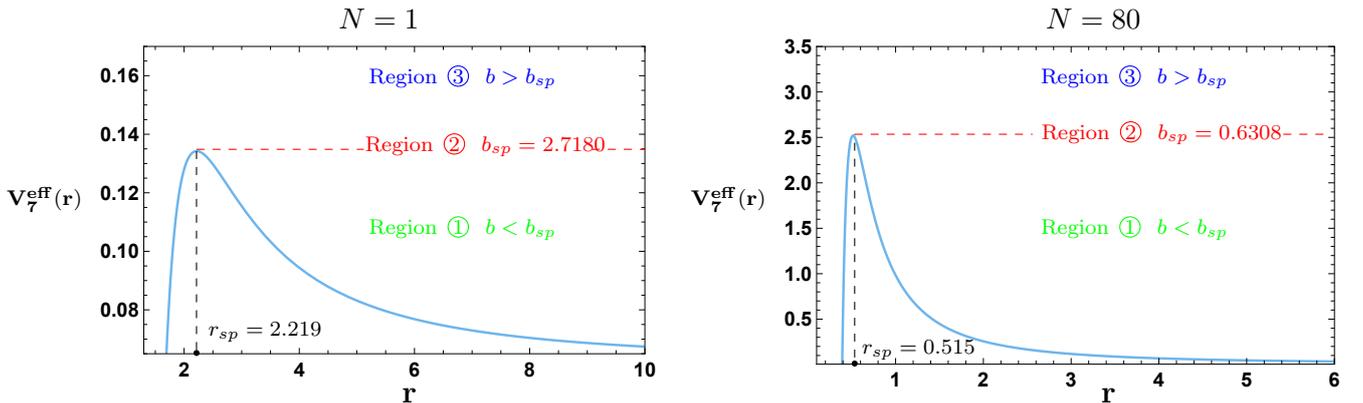

  Here,  we deal with the trajectory of the light rays near  black holes in the presence of  $M5$-branes in M-theory compactifications on the four-dimensional sphere  $S^4$.  For  the $(11,7,-3)$  model,  such light behaviors can be determined with the help of the effective potential
\begin{equation}
\label{ }
V^{eff}_{7}(r)=\frac{{L}^2}{r^2}\left(1+\frac{r^2}{4 \pi ^{2/3} N^{2/3} \ell _p^2}-\frac{6 \pi ^{5/3} M \ell _p^3}{5 N^{4/3} r^4}\right)-E^2
\end{equation}
In Fig.(\ref{aa2b}), we illustrate the associated  effective potential  as  a function of the  radial coordinate $r$ for two  values of the $M5$-brane number being  $N=1$ and $N=80$.
Using  the  constraint given by  Eq.(\ref{p1}), we can get the values of the photon sphere and the critical impact parameter of the photon sphere associated with the maximal value of the effective potential $V^{eff}_{7}(r).$
 
\begin{figure*}[!ht]
		\begin{center}
		\begin{tikzpicture}[scale=0.2,text centered]
		\hspace{-0.5 cm}
\node[] at (-30,1){\small  \includegraphics[scale=0.40]{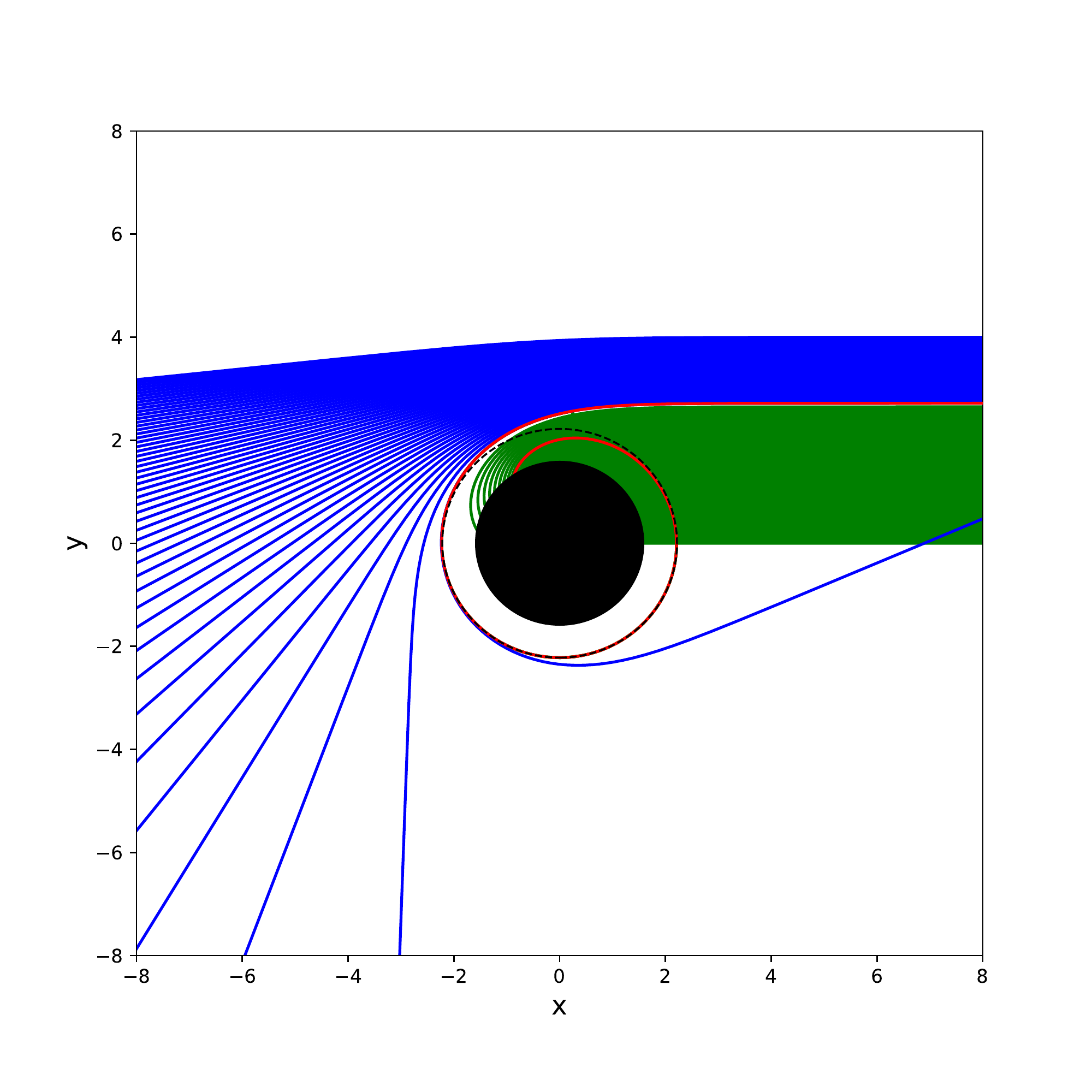}};
\node[] at (10,1){\small  \includegraphics[scale=0.40]{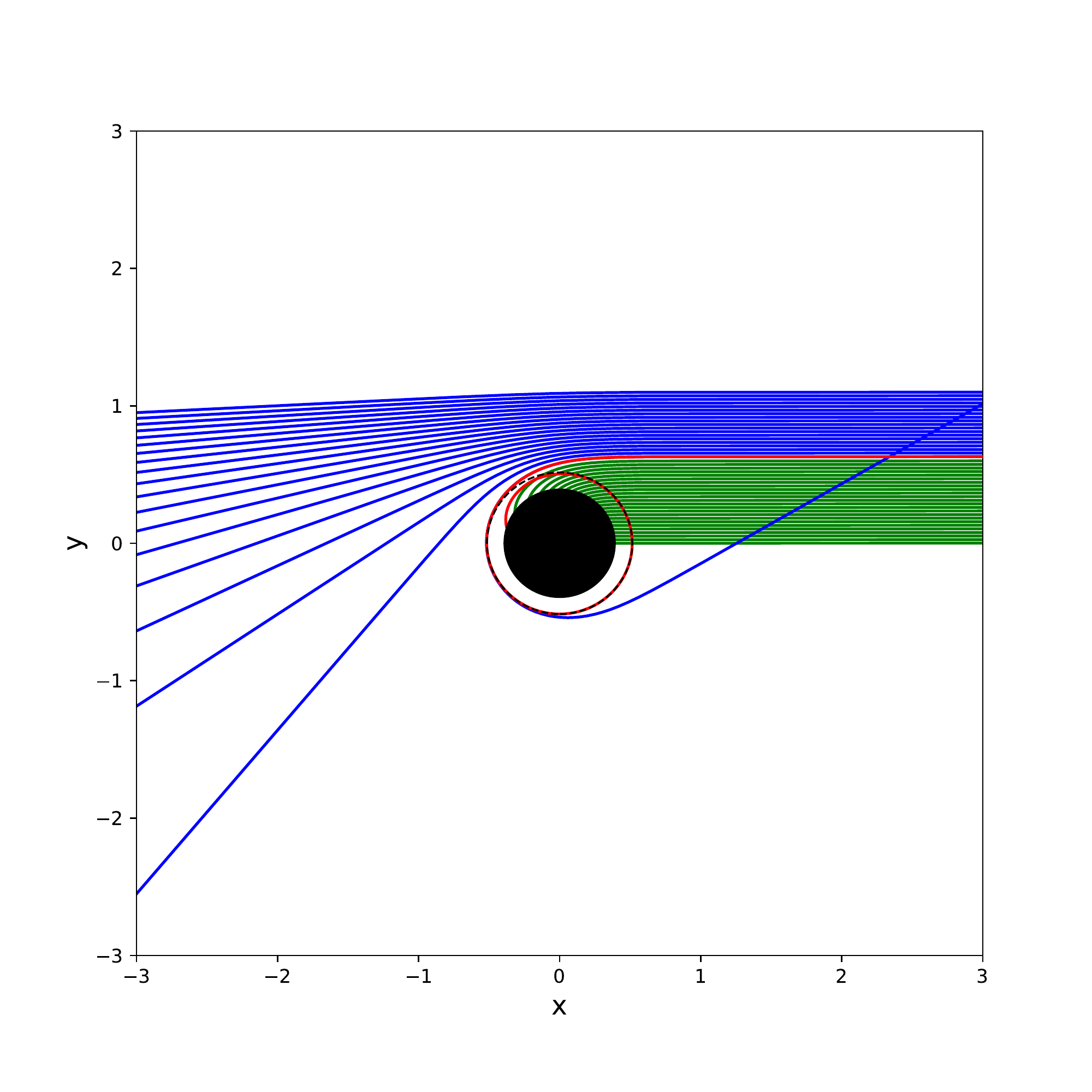}};
\node[color=black] at (-27.8,18.5) {$N=1$};
\node[color=black] at (12.8,18.5) {$N=80$};
\end{tikzpicture}	
\caption{ \footnotesize The trajectories of the light ray for different values of  $M5$-brane number. The black and the dashed red circle is the horizon and the photon sphere of the $M5$-brane, respectively.}
\label{aa1}
\end{center}
\end{figure*}

It follows  that  the two  values of  the $M5$-brane number $N=1$ and $N=80$ provide two  impact parameter values $b_{sp}=2.7180$ and $b_{sp}= 0.6308$ associated with  the  values of  the photon sphere $r_{sp}=2.219$ and $r_{sp}=0.515$, respectively. The corresponding impact parameter and the photon sphere radius decrease by increasing the $M5$-brane number. For the fixed value of $N=80$, however,  we observe that  $b_{sp}$ and $r_{sp}$ in the  $M5$-brane model  is small  compared  to  the  $M2$-brane model. This distinction affects the light ray trajectories around the  black holes in the  $M5$-brane model.  Placing  the observer in the equatorial hyperplane ensured by 
$\theta_1=\theta_2= \frac{\pi}{2}$,   we plot in Fig. (\ref{aa1}) the trajectories of the light rays in the polar coordinates ($r$,$\phi$) for different values of  the $M5$-brane number.

 Due to small values of  $b_{sp}$,   we vary the impact parameter $b$ by using   the step between  two values of   the impact parameter as 1/40 for all light rays.  It has been suggested that  the $M5$-brane number and the extra dimension can be considered as  relevant parameters  modifying    the  light ray behaviors  near a black hole in M-theory compactifications.  Such modifications come from the black hole geometry in the presence of the  $M5$-branes in M-theory compacatifications. The obtained behaviors match perfectly with   the previous works\cite{ma,B12}.

It has been remarked that the previous behaviors  of  the $M2$-brane model associated with  the horizon and  the  photon sphere  radius  have been  conserved in  the $M5$-brane model.  Concretely, one can show that the reflected light ray      becomes  more intense by deceasing the $M5$-brane number.  It has been remarked that these behaviors  are  contrary to the ones of the $M2$-brane  model. This is due to the  extra dimension contributions of the  $M5$-brane model  appearing  in  the associated metric function. Moreover,      $b_{sp}$ and $r_{sp}$    vary  by  taking different values of    the $M5$-brane number. For  all values of the  impact parameter,  we  observe  that if the light ray falls  into the black hole  or refracted   it keeps a  parallel trajectory with respect to  the $r$-axis.  This behavior is completely different than the  previous results related to  the $M2$-brane model.  Moreover,  the values of the angular velocity in  the $M5$-brane model  is different  than  the  one in the  $M2$-brane model.  A close examination shows that the results concerning the light trajectories behaviors confirm  the results associated with  the deflection angle  behavior.   It has been remarked that the deflection angle increases by decreasing the $M$-theory brane number $N$. Moreover,  the deflection of the light ray is more intense by decreasing the $M$-brane number. Both results inter-match  perfectly. Finally, the behaviors of the optical quantities including the light trajectories and the deflection angle of black holes for  M-theory  brane models are interesting and similar.
\section{Conclusion}
In this work, we have  investigated the deflection angle and the trajectory of  the light rays  casted by black holes in M-theory scenarios. Using the Gauss-Bonnet theorem, we  have  computed and examined  the deflection angle of the light rays around   four and seven dimensional AdS black holes derived  from the M-theory compacatification  on the real spheres  $S^7$  and $S^4$, respectively.  First, we   have generalized the deflection angle formalism for $d$-dimensional  AdS black hole solutions using the Gauss-Bonnet theorem.
Then,  we  have studied  the deflection angle of four dimensional rotating and non-rotating AdS black holes  by examining the $M2$-brane number effect for both  cases.   Concretely,  we have  shown  that  the  deflection angle  of light rays decreases for small value of the impact parameter then it becomes an increasing function.  It has been observed that  the  M-theory brane number decreases  the deflection angle.  Taking 
  two values of the impact parameter and varying $N$, we have observed that  the two curves meets a particular point, where the deflection angle of the AdS space has  changed   the behavior from a decreasing   to an increasing  function  in terms  the impact parameter. For   four dimensional rotating model, we  have revealed  that the behaviors of the deflection angle by varying  the  $M2$-brane number is  similar to the  non-rotating case. For the small  value of $a$, however,  the deflection angle of  the light rays  decreases for small values of $b$ and then it becomes an increasing  function of the impact parameter $b$. In general,  
 the rotating parameter  is  a relevant quantity  decreasing the deflection angle. We have shown that,   around $b=2$, the deflection angle behaviors depend on the $M2$-brane number. For generic values of $a$, the deflection angle increases by decreasing $N$. Similar behaviors  have been obtained in the previous results.  The only difference  around $b=2$ is the linear behavior   for large values of the  $M2$-brane number. 

Then, we   have   extended  the calculations to  seven dimensional non-rotating black holes  by providing   a comparative study. 
For small values of the impact parameter, we  have  shown  that the four-dimensional black hole bends the light rays larger than the seven-dimensional one.  However, this behavior is inverted for large values of  the impact parameter $b$.

Finally, we have  discussed  the trajectories  of the lights rays around four and seven-dimensional AdS black  hole in M-theory.  We  have shown  that the regions \let\textcircled=\pgftextcircled\textcircled{1}, \let\textcircled=\pgftextcircled\textcircled{2} and \let\textcircled=\pgftextcircled\textcircled{3}  are  the same for different  values of  the $M2$-brane number.  Taking two values  of $N$,  we have observed  that the  distance  between  two   light rays increases  for  an impact parameter  value  closer and bigger to  $b_{sp}$ value for all regions.  This  distance decreases  by increasing $N$. Concretely,  this distinction is originated  from the values of the    angular velocity.    We have remarked   that the  reflected  of  the light ray is more intense by decreasing the  $M2$-brane number.
Comparing  the present results   with many  investigations  associated with  the trivial solution, we have  observed  a different behavior. For the  $M5$-brane model, we  have remarked  a different behavior  with respect to the  $M2$-brane model. This difference could come  from the extra dimension being a relevant parameter appearing in the metric function.  \\ \\
This work comes up with certain open questions.  A possible project concerns generic models associated with   $(D,d,k)$   M-theory inspired models proposed in \cite{ma}.  In particular,  the effect of the  M$(d-2)$-branes in such models could be examined by performing  non trivial numerical computations.     Another issue is to approach certain  M-theory compactifications using  the orbifold of spheres providing possible ways to implement $G2$-manifolds in such black hole activities.   We hope address  these  questions in future works. 
\section*{Acknowledgments} 
 The present paper  is dedicated to the memory of  Pr. Ahmed  Intissar.  This work is partially
supported by the ICTP through AF.

\end{document}